\documentclass{article}

\pdfoutput=1
\usepackage{arxiv}
\usepackage{setspace}

\usepackage[utf8]{inputenc} 
\usepackage[T1]{fontenc}    
\usepackage{hyperref}       
\usepackage{url}            
\usepackage{booktabs}       
\usepackage{amsfonts}       
\usepackage{nicefrac}       
\usepackage{microtype}      
\usepackage{lipsum}
\usepackage{graphicx}
\usepackage{amsmath}
\usepackage{multirow}
\usepackage{array} 
\graphicspath{ {./images/} }

\makeatletter
\let\@fnsymbol\@arabic
\makeatother

\title{Advances in Land Surface Model-based Forecasting: A comparative study of LSTM, Gradient Boosting, and Feedforward Neural Network Models as prognostic state emulators}

\author{
Marieke Wesselkamp \thanks{Department of Biometry, University of Freiburg, Germany} \\
   \And
Matthew Chantry \thanks{European Centre for Medium-Range Weather Forecasts, Reading, United Kingdom} \\
   \And 
Ewan Pinnington \footnotemark[2] \\
   \And 
Margarita Choulga \footnotemark[2]  \\
   \And 
Souhail Boussetta  \footnotemark[2]  \\
   \And 
Maria Kalweit \thanks{Department of Computer Science, University of Freiburg, Germany}\\
   \And 
Joschka Boedecker \textsuperscript{\footnotemark[3]\hspace{0.7em}\footnotemark[4]}\\
   \And 
Carsten F. Dormann \footnotemark[1] \\
   \And
Florian Pappenberger  \footnotemark[2]  \\
   \And
Gianpaolo Balsamo\textsuperscript{\footnotemark[2]\hspace{0.7em}\footnotemark[5]}\\
}

\begin{document}
\maketitle

\footnotetext[4]{BrainLinks-BrainTools, Freiburg, Germany}
\footnotetext[5]{World Meteorological Organization, Geneva, Switzerland}
\begin{abstract}
Most useful weather prediction for the public is near the surface. The processes that are most relevant for near-surface weather prediction are also those that are most interactive and exhibit positive feedback or have key role in energy partitioning. Land surface models (LSMs) consider these processes together with surface heterogeneity and forecast water, carbon and energy fluxes, and coupled with an atmospheric model provide boundary and initial conditions. This numerical parametrization of atmospheric boundaries being computationally expensive, statistical surrogate models are increasingly used to accelerated progress in experimental research. We evaluated the efficiency of three surrogate models in speeding up experimental research by simulating land surface processes, which are integral to forecasting water, carbon, and energy fluxes in coupled atmospheric models. Specifically, we compared the performance of a Long-Short Term Memory (LSTM) encoder-decoder network, extreme gradient boosting, and a feed-forward neural network within a physics-informed multi-objective framework. This framework emulates key states of the ECMWF's Integrated Forecasting System (IFS) land surface scheme, ECLand, across continental and global scales. Our findings indicate that while all models on average demonstrate high accuracy over the forecast period, the LSTM network excels in continental long-range predictions when carefully tuned, the XGB scores consistently high across tasks and the MLP provides an excellent implementation-time-accuracy trade-off. The runtime reduction achieved by the emulators in comparison to the full numerical models are significant, offering a faster, yet reliable alternative for conducting numerical experiments on land surfaces.
\end{abstract}


\section{Introduction}

While forecasting of climate and weather system processes has long been a task for numerical models, the recent development in deep learning has introduced competitive machine-learning (ML) systems for numerical weather prediction (NWP) \cite{lamLearningSkillfulMediumrange2023a, biPanguWeather3DHighResolution2022, langAIFSECMWFDatadriven2024}. Land surface models (LSMs), even though being an integral part of numerical weather prediction, have not yet caught the attention of the ML-community. LSMs forecast water, carbon and energy fluxes, and in coupling with an atmospheric model, provide the lower boundary and initial conditions \cite{boussettaECLandECMWFLand2021, derosnayInitialisationLandSurface2014}. The parametrization of land surface states thus does not only affect predictability of earth and climate systems on sub-seasonal scales \cite{munoz-sabaterERA5LandStateoftheartGlobal2021}, but also the short- and medium-range skill of NWP forecasts \cite{derosnayInitialisationLandSurface2014}. Beyond the online integration with NWPs, offline versions of LSMs provide research tools for experiments on the land surface \cite{boussettaECLandECMWFLand2021}, the diversity of which are however limited by the required substantial computational resources and often moderate runtime efficiencies \cite{reichsteinDeepLearningProcess2019}. 

Emulators constitute statistical surrogates for numerical simulation models that, by approximating the latter, aim at increasing computational efficiency \cite{machacEmulationDynamicSimulators2016}. While for construction  emulators can themselves require substantial computational resources, their subsequent evaluation usually runs orders of magnitude faster than the original numerical model \cite{ferLinkingBigModels2018}. For this reason, emulators  have found application for example in modular parametrization of online weather forecasting systems \cite{chantryMachineLearningEmulation2021}, in replacing the MCMC-sampling procedure in Bayesian calibration of ecosystem models \cite{ferLinkingBigModels2018}, or in generating ensembles of atmospheric states for forecast uncertainty quantification \cite{liSEEDSEmulationWeather2023}. Beyond their computational efficiency, surrogate models with high parametric flexibility have the potential to correct for process mis-specification and improve predictions towards a physical model \cite{wesselkampProcessguidanceImprovesPredictive2022}. 

Modelling approaches used for emulation range from low parametrized, auto-regressive linear models to highly non-linear and flexible neural networks \cite{nathMESMERMEarthSystem2022, bakerEmulationHighresolutionLand2022a, meyerMachineLearningEmulation2022}. In the global land surface system M-MESMER, a set of simple AR1 regression models is used to initialize the numerical LSM, resulting in a modularized emulator \cite{nathMESMERMEarthSystem2022}. Numerical forecasts of gross primary productivity and hydrological targets were successfully approximated by Gaussian processes \cite{machacEmulationDynamicSimulators2016, bakerEmulationHighresolutionLand2022a}, the advantage of which is their direct quantification of prediction uncertainty. When it comes to highly diverse or structured data, neural networks have shown to deliver accurate approximations for variables from gravity wave drags to urban surface temperature \cite{chantryMachineLearningEmulation2021, meyerMachineLearningEmulation2022}. In most fields of machine learning, specific types of neural networks are now the best approach to representing fit and prediction. One exception is so-called tabular data, i.e. data without spatial or temporal interdependencies (as opposed to vision and sound), where extreme gradient boosting is still the go-to approach \cite{shwartz-zivTabularDataDeep2021, grinsztajnWhyTreebasedModels2022}. 

ECLand is the land surface scheme that provides boundary and initial conditions for the Integrated Forecasting System (IFS) of the European Centre for Medium-range Weather Forecasts (ECMWF) \cite{boussettaECLandECMWFLand2021}. Driven by meteorological forcing and spatial climate fields, it has a strong influence on the NWP \cite{derosnayInitialisationLandSurface2014} and also constitutes a standalone framework for offline forecasting of land surface processes, the advantage of which for the online framework is the temporal consistency of prognostic state variables \cite{munoz-sabaterERA5LandStateoftheartGlobal2021}. The modular construction of ECLand offers potential for element-wise improvement of process representation and thus a stepwise development towards increased computational efficiency. Within the IFS, ECLand also forms the basis of the land surface data assimilation system, updating the land surface state with synoptic data and satellite observations of soil moisture and snow. Emulators of physical systems have been shown to be beneficial in data assimilation routines, allowing for a quick and low maintenance estimation of the tangent linear model \cite{hatfieldBuildingTangentLinear2021, derosnayInitialisationLandSurface2014}. Together with the potential to run large ensembles of land surface states at a much-reduced cost, this would be a potential application of the surrogate models introduced here.
Long-short term memory networks (LSTMs) have gained popularity in hydrological forecasting as rainfall-runoff models, for predicting stream flow temperature and also soil moisture [e.g. \cite{kratzertLearningUniversalRegional2019, leesHydrologicalConceptFormation2022, zwartTermForecastsStream2023a, bassiLearningLandscapeFeatures2024}]. Research on the interpretability of LSTMs has found correlations between the model cell states and spatially or thematically similar hydrological units \cite{leesHydrologicalConceptFormation2022}, suggesting the specific usefulness of LSTM for representing variables with dynamic storages and reservoirs \cite{kratzertNeuralHydrologyInterpretingLSTMs2019}. As emulators, LSTMs have been shown useful for sea surface level projection in a variational manner with Monte Carlo dropout \cite{vankatwykVariationalLSTMEmulator2023}. While most of these studies trained their models on observations or reanalysis data, our emulator learns the representation from ECLand simulations directly. To our knowledge, a comparison of models without memory mechanisms to an LSTM-based neural network for global land surface emulation has not been conducted before. 

We emulate seven prognostic state variables of ECLand, which represent core land surface processes: soil water volume and soil temperature, each at three depth layers, and snow cover fraction at the surface layer. These three state variables represent the core of the current configuration of ECLand We specifically focus on the utility of memory mechanisms, highlighting the development of a single LSTM-based encoder-decoder model compared to an extreme gradient boosting approach (XGB) and a multilayer perceptron (MLP), which all perform the same tasks. The LSTM architecture builds on an encoder-decoder network design introduced for flood forecasting \cite{nearingGlobalPredictionExtreme2024}. To compare forecast skill systematically, the three emulators were compared in short- and long-range forecasting against climatology \cite{pappenbergerHowKnowIf2015}. In this work, evaluation is done on ECLand simulations only, i.e. on purely synthetic data, while future work will encompass transfer learning and validation on observations.

\section{Methods}
\label{sec:methods}

\subsection{The Land Surface Model: ECLand}

ECLand is a tiled ECMWF Scheme for Surface Exchanges over Land that represents surface heterogeneity and incorporates land surface hydrology (ECLand) \cite{balsamoEvolutionLandsurfaceProcesses2011, ecmwfIFSDocumentationCY43R32017}. ECLand computes surface turbulent fluxes (of heat, moisture and momentum) and skin temperature over different tiles (vegetation, bare soil, snow, interception and water) and then calculates an area-weighted average for the grid-box to couple with the atmosphere \cite{boussettaECLandECMWFLand2021}. For the overall accuracy of the model, accurate parameterizations are essential \cite{kimpsonDeepLearningQuality2023} as e.g. the land surface parameterization determines the sensible and latent heat fluxes, and provide the lower boundary conditions for enthalpy and moisture equations in the atmosphere \cite{viterboLand_surface_processes2002}. We emulate three prognostic state variables of ECLand, that represent core land surface processes: soil water volume and soil temperature at each three depth layers (each at 0 – 7 cm, 7 – 21 cm and 21 – 72 cm) and snow cover fraction, aggregated at the surface layer, so below are some more details on these parametrisations. 

\subsection{Data sources}

As training data base, global simulation and reanalysis time series from 2010 to 2022 were compiled to zarr format at an aggregated 6-hourly temporal resolution. Simulations and climate fields were generated from ECMWFs development cycle CY49R2, ECland forced by ERA-5 meteorological reanalysis data \cite{hersbachERA5GlobalReanalysis2020}. 
There are three main sources of data used for creation of the data base: The first is a selection of surface physiographic fields from ERA5 \cite{hersbachERA5GlobalReanalysis2020} and their updated versions \cite{choulgaUpgradedGlobalMapping2019, boussettaECLandECMWFLand2021, munoz-sabaterERA5LandStateoftheartGlobal2021} used as static model input features (X). The second is a selection of atmospheric and surface model fields from ERA5, used as static and dynamic model input features (Y). The third is ECLand simulation results, constituting the model’s dynamic prognostic state variables (z) and hence model input and target features. A total of 41 static, seasonal and dynamical features were used to create the emulators, see Table \ref{tab:climate_fields} for an overview of input variables and details on the surface physiographic and atmospheric fields below.

\subsubsection{Surface physiographic fields}

Surface physiographic fields have gridded information of the Earth’s surface properties (e.g.~land use, vegetation type, and distribution) and represent surface heterogeneity in the ECLand of the IFS \cite{kimpsonDeepLearningQuality2023}. They are used to compute surface turbulent fluxes (of heat, moisture, and momentum) and skin temperature over different surfaces (vegetation, bare soil, snow, interception, and water) and then to calculate an area-weighted average for the grid box to couple with the atmosphere. To trigger all different parametrization schemes, the ECMWF model uses a set of physiographic fields that do not depend on initial condition of each forecast run or the forecast step. Most fields are constant; surface albedo is specified for 12 months to describe the seasonal cycle. Depending on the origin, initial data come at different resolutions and different projections and are then first converted to a regular latitude–longitude grid (EPSG:4326) at approx.~1km at Equator resolution and secondly to a required grid and resolution. Surface physiographic fields used in this work consist of orographic, land, water, vegetation, soil, albedo fields, see Table \ref{tab:climate_fields} for the full list of surface physiographic fields; for more details, see IFS documentation \cite{ecmwfIFSDocumentationCY48R12023}. 

\subsubsection{ERA5}

Climate reanalyses combine observations and modelling to provide calculated values of a range of climactic variables over time. ERA5 is the fifth-generation reanalysis from ECMWF. It is produced via 4D-Var data assimilation of the IFS cycle 41R2 coupled to a land surface model (ECLand, \cite{boussettaECLandECMWFLand2021}), which includes lake parametrization by Flake \cite{mironovParameterizationLakesNWP} and an ocean wave model (WAM). The resulting data product provides hourly values of climatic variables across the atmosphere, land, and ocean at a resolution of approximately 31km with 137 vertical sigma levels up to a height of 80km. Additionally, ERA5 provides associated uncertainties of the variables at a reduced 63km resolution via a 10-member ensemble of data assimilations. In this work, ERA5 hourly surface fields at approx.~31km resolution on the cubic octahedral reduced Gaussian grid (i.e.~Tco399) are used. The Gaussian grid’s spacing between latitude lines is not regular, but lines are symmetrical along the Equator; the number of points along each latitude line defines longitude lines, which start at longitude 0 and are equally spaced along the latitude line. In a reduced Gaussian grid, the number of points on each latitude line is chosen so that the local east–west grid length remains approximately constant for all latitudes (here, the Gaussian grid is N320, where N is the number of latitude lines between a pole and the Equator).

\begin{table}[h]
\label{tab:climate_fields}
\caption{Input and target features to all emulators from the data sources. The left column shows the observation-derived static physiographic fields, the middle column ERA5 dynamic physiographic and meteorological fields and the rightmost column ECLand generated dynamic prognostic state variables.}
\centering
\begin{tabular}{p{3.5cm} p{1cm} p{3.5cm} p{1cm} p{3.5cm} p{1cm}}
\toprule
\textbf{Climate fields} & \textbf{Units} & \textbf{Atmospheric forcing} & \textbf{Units} & \textbf{Prognostic states} & \textbf{Units}  \\
\midrule
Vegetation cover (low, high) &  & Total precipitation fraction (convective, stratiform) & & Soil water volume (Layers 1-3) &  m\(^3\)m\(^{-3}\) \\
Type of vegetation (low, high) &  & Downward radiation (long, short)  & W/m\(^2\) & Soil temperature (Layers 1-3) & K \\
Minimum stomatal resistance (low, high) &  & Seasonal LAI (high, low)& & Snow cover fraction & \\
Roughness length (low, high) &  & Wind speed (v, u)  & m/s & &\\
Urban cover &  & Surface pressure  & Pa & &\\
Lake cover &  & Skin temperature  & K & &\\
Lake depth &  &  & & &  \\
Orography (std, filtered)  & m\(^2\)/s\(^2\) & Specific humidity  & kg/kg & & \\
Photosynthesis pathways &  & Rainfall rate (total) & kg/m\(^2\)s & & \\
Soil type &  & Snowfall rate (total)  & kg/m\(^2\)s & &\\
Glacier mask &  &  & & & \\
Permanent wilting point &  &  & & &  \\
Field capacity &  &  & & &  \\
Cell area &  &  & & &  \\
\bottomrule
\end{tabular}
\end{table}

\subsection{Emulators}

We compare the utility of a long-short term memory neural network (LSTM), that of extreme gradient boosting regression trees (XGB) and that of a feedforward neural network (that we here refer to as multilayer perceptron, MLP). To motivate this setup and pave the way for discussing effects of (hyper-)parameter choices, a short overview of all approaches is given. All analyses were conducted in Python. XGB was developed in dmlc’s XGBoost python package\footnote{https://xgboost.readthedocs.io/en/stable/python/index.html}. The MLP and LSTM were developed in the PyTorch lightning framework for deep learning\footnote{https://lightning.ai/docs/pytorch/stable/}. Neural networks were trained with the Adam algorithm for stochastic optimization \cite{kingmaAdamMethodStochastic2017}. Model architectures and algorithmic hyperparameters were selected through Bayesian hyperparameter optimization with the Optuna framework \cite{akibaOptunaNextgenerationHyperparameter2019}. The Bayesian optimization minimizes the neural network validation accuracy, specified here as mean absolute error (MAE), over a predefined search space for free hyperparameters with the Tree-structured Parzen Estimator \cite{ozakiMultiobjectiveTreeStructuredParzen2022}. The resulting hyperparameter and architecture choices which were used for the different approaches are listed in the Appendix.

\subsubsection{MLP}

For creation of the MLP emulator we work with a feed-forward neural network architecture
of connected hidden layers with ReLU activations and dropout layers, model components which are given in detail in the Appendix or in \cite{goodfellowDeepLearning2016}. The MLP was trained with a learning rate scheduler. L2-regularization was added to the training objective via weight decay. Sizes and width of hidden layers as well as hyperparameters were selected together in the hyperparameter optimization procedure. Instead of forecasting absolute prognostic state variables $z_t$, the MLP predicts the 6-hourly increment, $\frac{\widehat{dz}}{dt}$. It is trained on a stepwise rollout prediction of future state variables at a pre-defined lead time at given forcing conditions, see details in the section on optimization.

\subsubsection{LSTM}

LSTMs are recurrent networks that consider long-term dependencies in time series through gated units with input and forget mechanisms \cite{hochreiterLongShortTermMemory1997}. In explicitly providing time-varying forcing and state variables, LSTM cell states serve as long-term memory while LSTM hidden states are the cells’ output and pass on stepwise short-term representations stepwise. In short notation \cite{leesHydrologicalConceptFormation2022}, a one-step ahead forward pass followed by a linear transformation can be formulated as
\begin{equation}
\begin{aligned}
    h_t,\ c_t & =f\left(x_t,\ h_{t-1},\ c_{t-1},\ \theta\right)
\ \\
 {\hat{z}}_t & =\ {Ah}_t+b
\end{aligned}
\end{equation}
where $h_{t-1}$ denotes the hidden state, i.e. output estimates from the previous time step, $c_{t-1}$ the cell state from the previous time step, and $\theta$ the time-invariant model weights. We stacked multiple LSTM cells to an encoder-decoder model with transfer layers for hidden and cell state initialization and for transfer to the context vector (see figure 1) \cite{nearingGlobalPredictionExtreme2024}. A lookback $l$ of the previous static and dynamic feature states are passed sequentially to the first LSTM cells in the encoder layer, while the l prognostic state variables $z$ initialize the hidden state $h_0$ after a linear embedding. The output of the first LSTM layer cells become the input to the deeper LSTM layer cells and the last hidden state estimates are the final output from the encoder. Followed by a non-linear transformation with hyperbolic tangent activation, the hidden cell states are transformed into a weighted context vector $s$. Together with the encoder the cell state ($c_t$, $s$) initializes the hidden and cell states of the decoder. The decoder LSTM cells take as input again static and dynamic features sequentially at lead times $t=1,\ \ldots,\tau$, but not the prognostic states variables. These are estimated from the sequential hidden states of the last LSTM layer cells, transformed to target size with a linear forecast head before prediction. LSTM predicts absolute state variables $z_t$ while being optimized on $z_t$ and ${d\hat{z}}_t$ simultaneously, see section on optimization.

\begin{figure}
    \centering
    \includegraphics[width=0.8\textwidth]{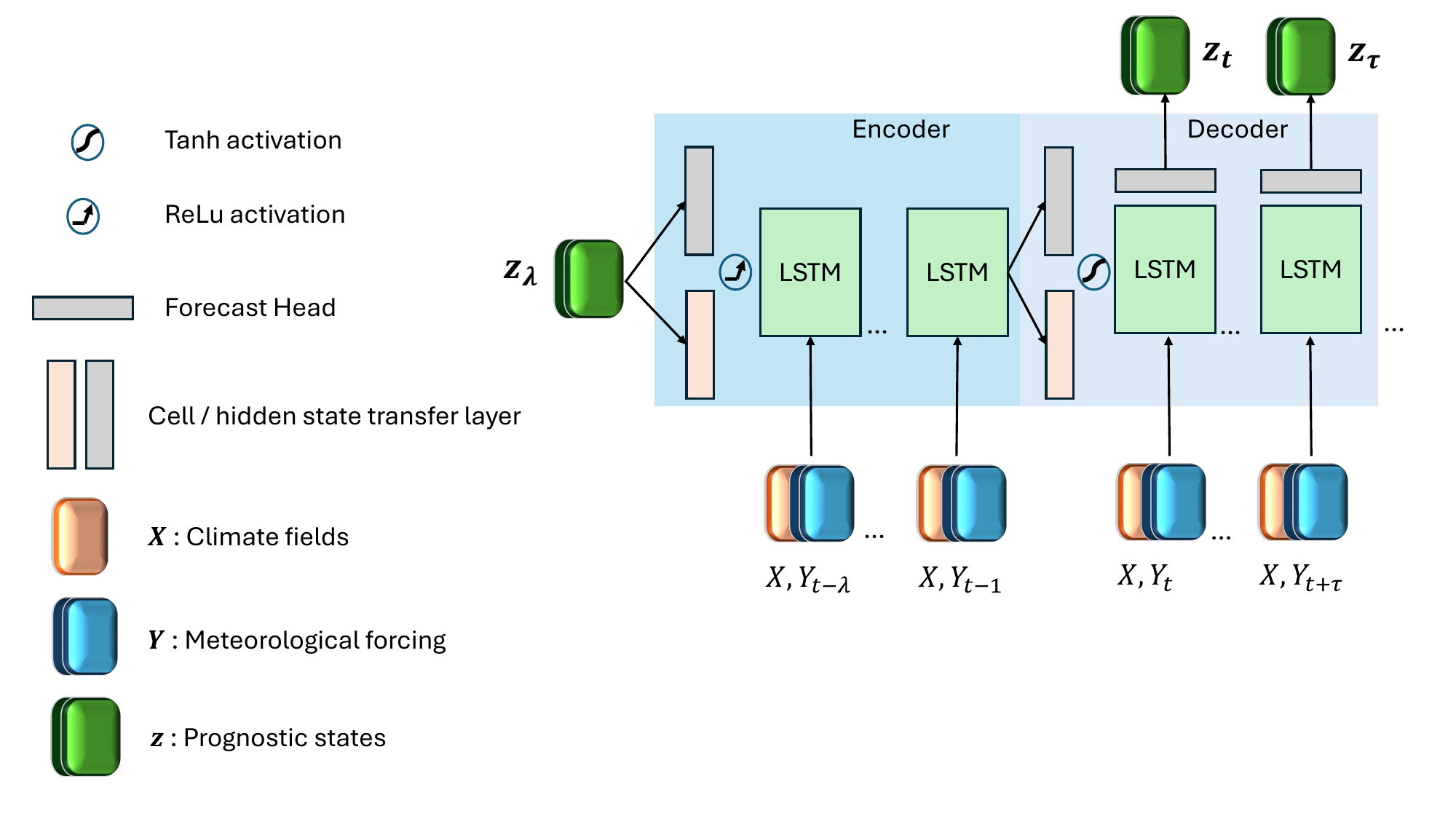}
    \caption{LSTM architecture. Blue shaded area indicates the encoder part, where the model is driven by a lookback $\lambda$ of meteorological forcing and state variables. The light-blue shaded area indicates the decoder part that is initialized from the encoding to unroll LSTM forecasts from the initial time step $t$ up to a flexibly long lead time of  $\tau$.}
    \label{fig:lstm}
\end{figure}

\subsection{Optimization}

\subsubsection{Loss functions}

The basis of the loss function $\mathcal{L}$ for the neural network optimization was PyTorch’s SmoothL1Loss , a robust loss function that combines L1-norm and L2-norm and is less sensitive to outliers than pure L1-norm \cite{girshickFastRCNN2015}. Based on a pre-defined threshold parameter $\beta$, smooth L1 transitions from L2-norm to L1-norm above the threshold. 
SmoothL1Loss $\mathcal{L}$ is defined as
\begin{equation}
\begin{aligned}
\mathcal{L}(\hat{z},\ \ z) & =\ 0.5\left(\hat{z}-\ z\right)^2\frac{1}{\beta}  if |\hat{z}-\ z|<\ \beta\ \text{and} \\
\mathcal{L}(\hat{z},\ \ z) & =\ |\hat{z}-\ z|-\ 0.5\ \beta  \text{otherwise},
\end{aligned}
\end{equation}

here with $\beta=1$. All models were trained to minimize the incremental loss $\mathcal{L}_s$ that is the differences between the estimates of the seven prognostic states increments  $\widehat{dz}_t$ and the full model’s prognostic states increments ${dz}_t$simultaneously as the sum of losses over all states. We opted for an loss function equally weighted by variables to share inductive biases among the non-independent prognostic states \cite{senerMultiTaskLearningMultiObjective2018}. When aggregating over all training lead times $t=1,\ \ldots,\ \tau$, $\mathcal{L}_s$ and grid cells $i=1,\ldots,p$ is 
\begin{equation}
\begin{aligned}
\mathcal{L}_s(\ \widehat{dz},\ dz)=\ \sum_{t}^{\tau}\sum_{i}^{p}{\mathcal{L}_t(\ {\widehat{dz}}_{t,i},\ {dz}_{t,i})},
\end{aligned}
\end{equation}
Whereas when computing a rollout loss $\mathcal{L}_r$ stepwise, 
\begin{equation}
\begin{aligned}
\mathcal{L}_r(\ \widehat{dz},\ z)=\ \frac{1}{\tau\ }\sum_{t}^{\tau}\sum_{i}^{p}{\mathcal{L}_t(z_{t-1,i}+\ {\widehat{dz}}_{t,i},\ z_{t,i})}
\end{aligned}
\end{equation}
Prognostic state increments are essentially the first differences from one to the next timestep that are normalized again by the global standard deviation of the model’s states increments, $s_{\ _{dz}}$ before computation of the loss \cite{keisler_forecasting_2022}. Due to the forecast models’ structural differences, loss functions were individually adapted.

\paragraph{MLP} The combined loss function for the MLP is the sum of the incremental loss $\mathcal{L}_s$ and the rollout loss $\mathcal{L}_r$. For the rollout loss $\mathcal{L}_r$, $\mathcal{L}$ was aggregated over grid cells p and accumulated after an auto-regressive rollout over lead times $\tau$, before being averaged out by division by $\tau$ \cite{keislerForecastingGlobalWeather2022}. 
\paragraph{LSTM} The combined loss function for the LSTM is the sum of the incremental loss $\mathcal{L}_s$, where the ${d\hat{z}}_t$ were derived from ${\hat{z}}_t$ after the forward pass, and the loss $\mathcal{L}$ computed on decoder estimates of prognostic states variables, a functionality that leverages the potential of our LSTM structure. 
\paragraph{XGB} Trained only from one to the next time step, i.e. at a lead time of $\tau=1$, the incremental loss $\mathcal{L}_s=\ \mathcal{L}_r$. Without a SmoothL1Loss implementation provided in dmlc’s XGBoost, we trained XGB with both the Huber-Loss and the default L2-loss. The latter initially providing better results, we chose the default L2-norm as loss function for XGB with the regularization parameter $\lambda=1$.

\subsubsection{Normalization}

As prognostic target variables are all lower bounded by zero, we tested both z-scoring and max-scoring. The latter yielded no significant improvement, thus we show our results with z-scored target variables. For neural network training but not for fitting XGB, static, dynamic and prognostic state variables were all normalized with z-scoring towards the continental or global mean $\bar{z}$ and unit standard deviation $s_z$ as
\begin{equation}
    z_{t,n}=\ \frac{z_{t,n}-\ \bar{z}}{s_z}.
\end{equation}
Prognostic target state increments were normalized again by the global standard deviation of increments computing the loss (see section 2.5.1) to smooth magnitudes of increments \cite{keislerForecastingGlobalWeather2022}. State variables were backtransformed to original scale before evaluation.

\subsubsection{Spatial and temporal sampling}
Sequences were sampled randomly from the training data set, while validation happened sequentially. MLP and XGB were trained on all grid cells simultaneously in both the continental and global setting, while LSTM was trained on the full continental data set but was limited by GPU memory in the global task. We overcame this limitation by randomly subsetting grid cells in the training data into largest possible, equally sized subsets which were then loaded along with the temporal sequences during the batch sampling. 

\subsubsection{Evaluation}

Three scores are used for model validation during the model development phase and in validating architecture and hyperparameter selection, being the root mean squared error (RMSE), the mean absolute error (MAE) and the anomaly correlation coefficient (ACC). First, scores were assessed objectively in quantifying forecast accuracy of the emulators against ECLand simulations directly with RMSE and MAE. Doing so, scores were aggregated over lead times, grid cells or both. The total RMSE was computed as 
\begin{equation}
\text{RMSE}=\ \sqrt{\frac{\sum_{\tau,\ p}{(z-\ \hat{z})}^2}{n}},
\end{equation}
As the mean absolute error in prognostic state variable prediction over the total of n grid cells p times lead times $\tau$. Equivalently, MAE was computed as
\begin{equation}
\text{MAE}=\ \frac{\sum_{t,p}{|z-\ \hat{z}|}}{n},
\end{equation}
Beyond accuracy, the forecast skill of emulators was assessed using climatology as a benchmark models. The ACC (see below) as index of the long-term naïve climatology c, i.e. the 6-hourly mean of prognostic state variables over the last 10 years preceding the test year, i.e. the years 2010 to 2020. While climatology is a hard-to-beat benchmark specifically in long-term forecasting, the persistence is a benchmark for short-term forecasting \cite{pappenberger_how_2015}. 
For verification against climatology, we compute the anomaly correlation coefficient (ACC) over lead times as 
\begin{equation}
\text{ACC}(t) = \frac{\overline{(\hat{z} - c)(z - c)}}{\sqrt{\overline{(\hat{z} - c)^2} \, \overline{(z - c)^2}}}
\end{equation}
at each $t = 1, …, \tau$ where the overbar denotes averaging over grid cells $p=i,\ \ldots,\ n$. The nominator now indicates the mean squared skill error towards climatology and the denominator its variability. ACC is bounded between 1 and -1, and an ACC of 1 indicates perfect representation of forecast error variability, an ACC of 0.5 indicates a similar forecast error to that of the climatology, an ACC of 0 indicates that forecast error variability dominates and the forecast has no value and an ACC approaching -1 indicates that the forecast has been very unreliable \cite{ecmwfForecastUserGuide}. ACC is undefined when the denominator is zero. This is the case either when mean squared emulator or ECLand anomaly, or both are zero because forecast and climatology perfectly align, or because they cancel out at summation to the mean.

\subsubsection{Forecast horizons}

Forecast horizons of the emulators are defined by the decomposition of the RMSE \cite{bengtssonIndependentEstimationsAsymptotic2008} into the emulator’s variability around climatology (i.e. anomaly), ECLand’s variability around climatology and the covariance of both. The horizon is the point in time at which the forecast error reaches saturation level, that is when the covariance of emulator and ECLand anomalies approaches zero, as does the ACC.
We analysed predictive ability and predictability by computing the ACC for all lead times from 6 hours to approx. one year, i.e. lead times $t=1,\ldots,\ \tau$, $\tau$ being 1350. As this confounds the seasonality with the lead time, we compute these for every starting point of the prediction, requiring two test years (2021 and 2022). 
Forecast horizons based on the emulators’ skill in standardized anomaly towards persistence were equivalently computed but with persistence as a benchmark for shorter time scales, this was only done for three months, from January to March 2021.
The analysis was conducted on two exemplary regions in northern and southern Europe that represent very different conditions orography and in prognostic land surface states, specifically in snow cover. For details on the regions and on the horizons computed with standardized anomaly skill, see Appendices A1 and A4 respectively.  

\section{Results}

\subsection{Aggregated performances}

\paragraph{Europe} All emulators approximated the numerical LSM with high average total accuracies (all RMSEs < 1.58 and MAEs < 0.84) and confident correlations (all ACC > 0.72) (see table 2 and figure 2). The LSTM emulator achieved the best results across all total average scores on the European scale. It decreased the total average MAE by $\sim$25\% towards XGB and by $\sim$37\% towards the MLP and the total average RMSE by $\sim$42\% towards XGB and $\sim$38\% towards the MLP.  In total average ACC, the LSTM scored 20\% higher than the MLP and 15\% than XGB, also being the only emulator that achieved an ACC > 0.9. While the MLP outperforms XGB in total average RMSE by $\sim$5\%, XGB scores better than the MLP in MAE by $\sim$27\%.
At variable level, results differentiate into model specific strengths. In soil water volume, XGB outperforms the neural network emulators by up to 60\% in the first and second layer MAEs towards the LSTM and up to over 40\% for towards the MLP (see table 3). While the representation of anomalies by specifically the LSTM decreases towards lower soil layers with an ACC of only 0.6214 at the third soil layer, it remains consistently higher for XGB with an ACC still > 0.789 at soil layer three. 

In soil temperature approximation, LSTM achieves best accuracies at higher soil levels with up to 7\% improvement in MAE towards XGB and ACCs > 0.92, but XGB outperforms LSTM at the third soil level with a close to 50\% improvement (see table 4). The MLP doesn’t stand out by high scores on the continental scale. However, in terms of accuracy we found an inverse ranking in the model development procedure during which LSTM outscored XGB in soil water volume but struggled with soil temperature approximations, for the interested reader we refer to the supplementary information.
In snow cover approximation, the LSTM emulator enhances accuracies by over $\sim$50\% in MAE towards both the XGB and the MLP emulator and scores highest in anomaly representation with an ACC of $\sim$0.87 compared to an ACC of $\sim$0.66 for the MLP and only $\sim$0.74 for the XGB (see table 5).

\paragraph{Globe} Score ranking on the global scale varies strongly from the continental scale (see table 2). In total average accuracies, the MLP outperforms XGB by over 30\% and LSTM by up to 25\% in RMSE and improves MAE more than 15\% towards both. In anomaly correlation however it scores last, whereas XGB achieves the highest total average of over 0.75. Consistent with scores on the continental scale is XGBs high performance in soil temperature (see table 3). It significantly outperforms the LSTM by $\sim$60\% in RMSE and nearly up to 75\% in MAE in all layers and the MLP by up to 50\% in MAE at the top layer. Anomaly persistence for all models degrade visibly towards the lower soil layers, while that of the LSTM most relative to MLP and XGB. Similar to the continental scale, XGB also outperforms the other candidates in soil temperature forecasts in all but the medium layer, where the MLP gets higher scores in MAE and RMSE but not in ACC (see table 4). LSTM doesn’t stand out with any scores on the global scale.

\begin{table}[h]
    \centering
    \caption{Emulator total average scores, aggregated over variables, time and space from the European and Global model testings}
    \label{tab:total_scores}
    \begin{tabular}{l | l | ll | ll | ll}
    \toprule
    \multirow{2}{*}{Variable} & \multirow{2}{*}{Model} & \multicolumn{2}{c|}{RMSE} & \multicolumn{2}{c|}{MAE} & \multicolumn{2}{c}{ACC} \\ 
                                  &                       & Europe & Globe & Europe & Globe & Europe & Globe \\ \hline
    \multirow{3}{*}{All variables} & XGB & 1.575 & 2.611 & 0.695 & 1.601 & 0.765 & 0.755 \\ 
                                       & MLP & 1.486 & \textbf{1.699} & 0.832 & \textbf{1.189} & 0.728 & \textbf{0.569} \\ 
                                       & LSTM & \textbf{0.918} & 2.252 & \textbf{0.526} & 1.787 & \textbf{0.925} & 0.647 \\ \hline
    \end{tabular}
\end{table}

\begin{table}[h]
\centering
\label{tab:swvl}
\caption{Emulator average scores on soil water volume forecasts for the European subset, aggregated over space and time from the European and Global model testing.}
\begin{tabular}{ll | l | lll | lll | lll}
\toprule
\textbf{Variable} & \textbf{Layer} & \textbf{Model} & \multicolumn{2}{c}{\textbf{RMSE}} & & \multicolumn{2}{c}{\textbf{MAE}} & & \multicolumn{2}{c}{\textbf{ACC}} \\
 & & & \textbf{Europe} & \textbf{Globe} & & \textbf{Europe} & \textbf{Globe} & & \textbf{Europe} & \textbf{Globe} \\
\midrule
\multirow{1}{*}{Soil water volume} & \multirow{0}{*}{1} & XGB & \textbf{0.013} & \textbf{0.015} & & \textbf{0.010} & \textbf{0.010} & & \textbf{0.908} & \textbf{0.920} \\
 & & MLP & 0.019 & 0.029 & & 0.015 & 0.023 & & 0.856 & 0.791 \\
 & & LSTM & 0.029 & 0.048 & & 0.023 & 0.040 & & 0.847 & 0.729 \\
 \midrule
& \multirow{0}{*}{2} & XGB & \textbf{0.011} & \textbf{0.012} & & \textbf{0.008} & \textbf{0.009} & & \textbf{0.901} & \textbf{0.884} \\
& & MLP & 0.019 & 0.023 & & 0.014 & 0.018 & & 0.789 & 0.770 \\
& & LSTM & 0.029 & 0.050 & & 0.023 & 0.042 & & 0.790 & 0.617 \\
\midrule
& \multirow{3}{*}{3} & XGB & \textbf{0.015} & \textbf{0.014} & & \textbf{0.011} & \textbf{0.010} & & \textbf{0.789} & \textbf{0.777} \\
& & MLP & 0.020 & 0.020 & & 0.017 & 0.016 & & 0.576 & 0.667 \\
& & LSTM & 0.033 & 0.051 & & 0.027 & 0.043 & & 0.621 & 0.475 \\
\bottomrule
\end{tabular}
\label{tab:performance_swvl}
\end{table}

\begin{table}[h]
\centering
\label{tab:stl}
\caption{Emulator average scores on soil temperature forecasts for the European subset, aggregated over space and time from the European and Global model testing.}
\begin{tabular}{ll | l | lll | lll | lll}
\toprule
\textbf{Variable} & \textbf{Layer} & \textbf{Model} & \multicolumn{2}{c}{\textbf{RMSE}} & & \multicolumn{2}{c}{\textbf{MAE}} & & \multicolumn{2}{c}{\textbf{ACC}} \\
 & & & \textbf{Europe} & \textbf{Globe} & & \textbf{Europe} & \textbf{Globe} & & \textbf{Europe} & \textbf{Globe} \\
\midrule
\multirow{1}{*}{Soil temperature} & \multirow{0}{*}{1} & XGB & 1.154 & 4.539 & & 0.744 & 3.278 & & 0.806 & \textbf{0.769} \\
 & & MLP & 1.628 & \textbf{2.606} & & 1.188 & \textbf{2.072} & & 0.674 & 0.581 \\
 & & LSTM & \textbf{0.931} & 3.152 & & \textbf{0.682} & 2.626 & & \textbf{0.938} & 0.735 \\
 \midrule
& \multirow{0}{*}{2} & XGB & 0.901 & 2.501 & & \textbf{0.51} & 1.772 & & 0.812 & \textbf{0.797} \\
& & MLP & 1.134 & \textbf{1.851} & & 0.784 & \textbf{1.452} & & 0.718 & 0.606 \\
& & LSTM & \textbf{0.734} & 2.87 & & 0.541 & 2.4 & & \textbf{0.928} & 0.699 \\
\midrule
& \multirow{3}{*}{3} & XGB & \textbf{0.714} & \textbf{1.287} & & \textbf{0.482} & \textbf{0.933} & & \textbf{0.722} & \textbf{0.711} \\
& & MLP & 1.128 & 1.375 & & 0.821 & 1.071 & & 0.416 & 0.514 \\
& & LSTM & 1.141 & 3.466 & & 0.918 & 3.002 & & 0.598 & 0.406 \\
\bottomrule
\end{tabular}
\label{tab:performance_stl}
\end{table}

\begin{table}[h]
\centering
\label{tab:snowc}
\caption{Emulator average scores on snow cover forecasts for the European subset, aggregated over space and time from the European and Global model testing.}
\begin{tabular}{ll | l | lll | lll | lll}
\toprule
\textbf{Variable} & \textbf{Layer} & \textbf{Model} & \multicolumn{2}{c}{\textbf{RMSE}} & & \multicolumn{2}{c}{\textbf{MAE}} & & \multicolumn{2}{c}{\textbf{ACC}} \\
 & & & \textbf{Europe} & \textbf{Globe} & & \textbf{Europe} & \textbf{Globe} & & \textbf{Europe} & \textbf{Globe} \\
\midrule
\multirow{1}{*}{Snow Cover} & \multirow{0}{*}{Top} & XGB & 8.219 & 9.906 & & 3.099 & 5.196 & & 0.746 & \textbf{0.707} \\
 & & MLP & 6.449 & \textbf{5.995} & & 2.986 & \textbf{3.671} & & 0.66 & 0.618 \\
 & & LSTM & \textbf{3.526} & 6.127 & & \textbf{1.47} & 4.357 & & \textbf{0.877} & 0.698 \\
\bottomrule
\end{tabular}
\label{tab:performance_snowc}
\end{table}

\subsection{Spatial and temporal performances}

\begin{figure}
    \centering
    \includegraphics[width=0.8\textwidth]{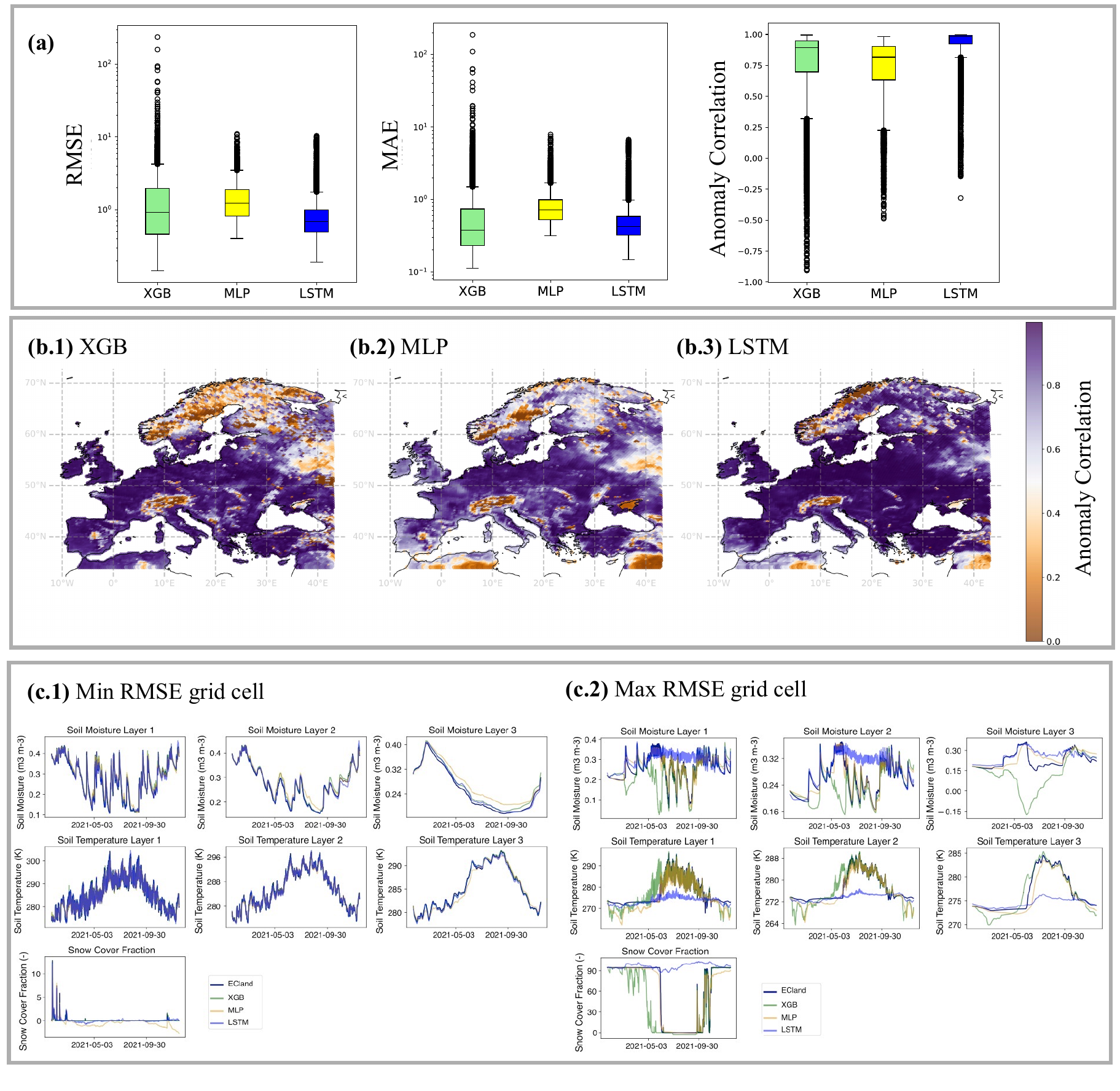}
    \caption{\textbf{a}: Total aggregated distributions of (log) scores averaged over lead times, i.e. displaying the variation among grid cells. \textbf{b}: The distribution of the anomaly correlation in space on the European subset (b.1: XGB, b.2: MLP, b.3: LSTM). \textbf{c}: Model forecasts over test year 2021 for grid cell with minimum and maximum RMSE values (LSTM).}
    \label{fig:europe}
\end{figure}

\begin{figure}
    \centering
    \includegraphics[width=0.75\textwidth]{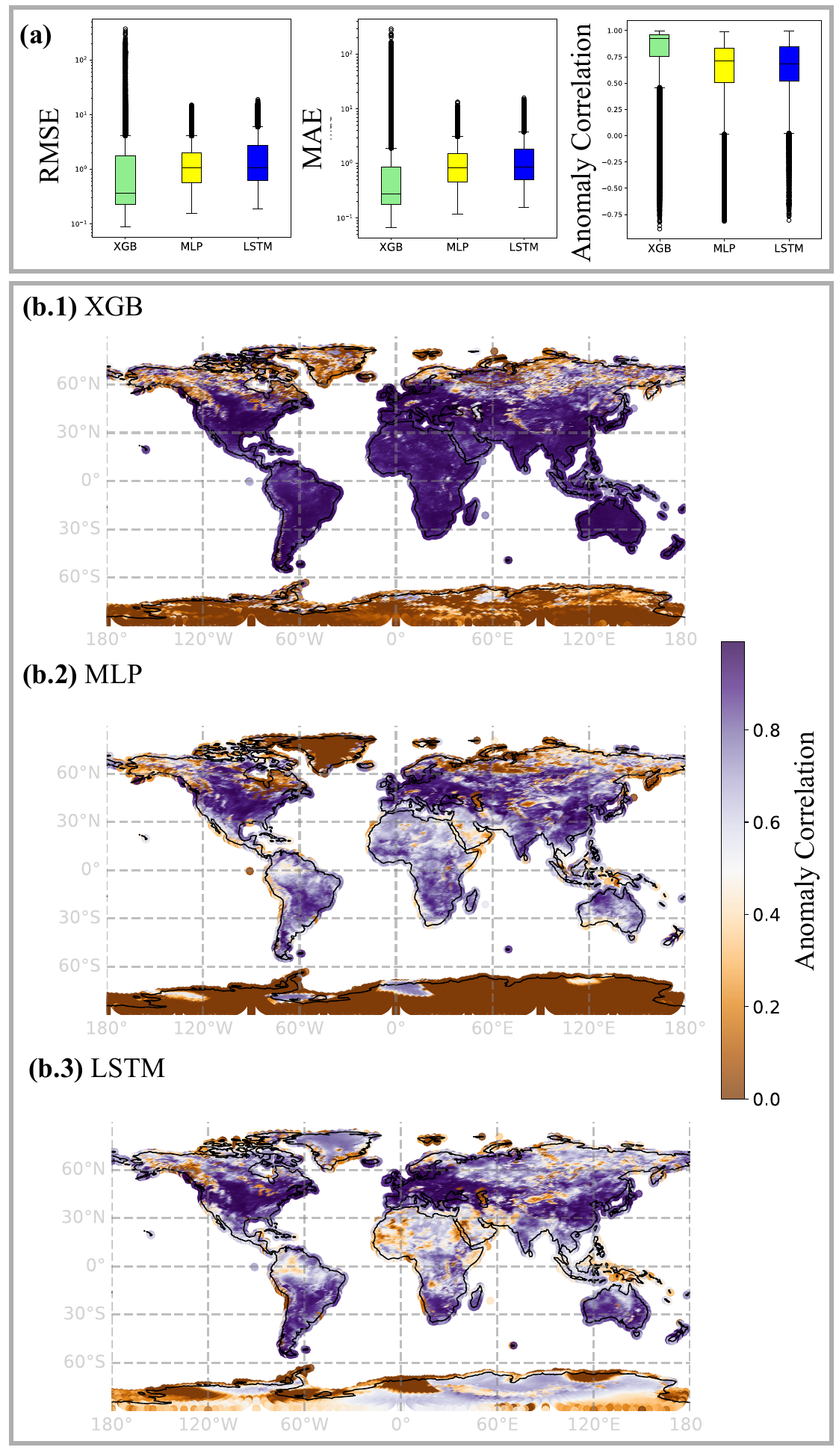}
    \caption{\textbf{a} Total average scores, representing spatial variation among grid cells. \textbf{b} Total average ACC in space. Note that ACC remained undefined for regions of low signal in snow cover and soil water volume, see Appendix.}
    \label{fig:globe}
\end{figure}

\paragraph{Europe} When summarizing temporally aggregated scores as boxplots to a total distribution over space (see Figure \ref{fig:europe}, a), the long tails of XGB scores become visible, whereas the MLP indicates most robustness. This is reflected in the geographic distribution of scores at the example of ACC (see Figure \ref{fig:europe}, bottom), where the area of low anomaly correlation is largest for XGB, ranging over nearly all northern Scandinavia, while MLP and LSTM have smaller and more segregated areas of clearly low anomaly correlation. The LSTM shows a homogenously high ACCs over most of central Europe but the Alps, while also seems to be challenged in areas of relative to the central Europe extreme weather conditions at the Norwegian and Spanish coasts.
\paragraph{Globe} Similar to the results from the continental analysis, we find again long upper tails of outliers for XGB in total spatial distribution of accuracies, both in RMSE and MAE and only few outliers for MLP and LSTM (see Figure \ref{fig:globe}). The anomaly correlation distribution changed towards longer lower tails for MLP and LSTM and a shorter lower tail for XGB. We should, however, take the results of total average ACC with care as it remains largely undefined in regions without much noise in snow cover or soil water volume and globally represents mainly patterns of soil temperature.

\subsection{Forecast horizons}

\begin{figure}
    \centering
    \includegraphics[width=0.8\textwidth]{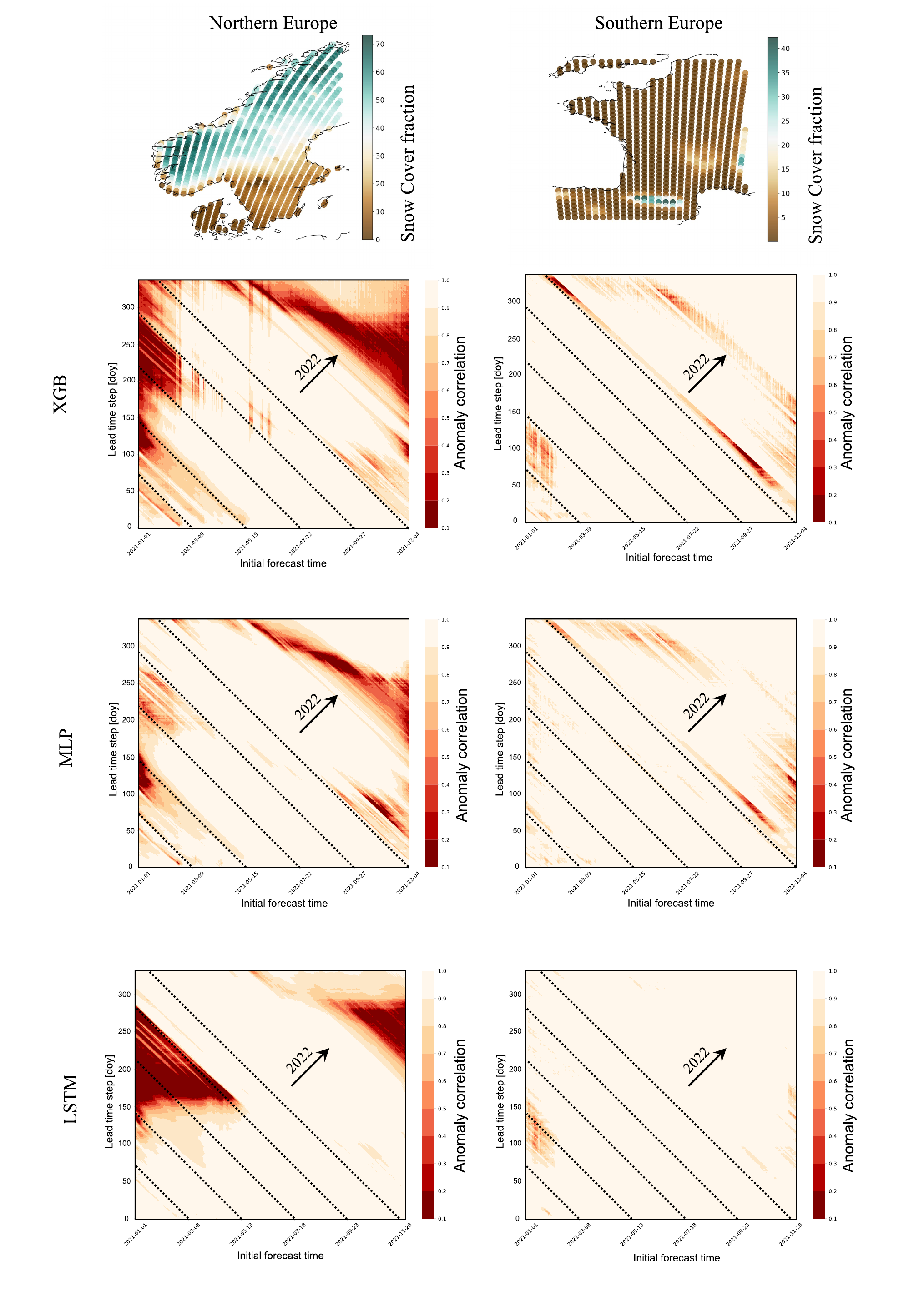}
    \caption{Emulator forecast skill horizons in two European subregions, aggregated over prognostic state variables. Scores are computed with the anomaly correlation coefficient (ACC) at 6-hourly lead times (y-axis) over approx. one year, displayed as a function of the initial forecast time (x-axis). As horizon we define the time at which the forecast has no value at all, i.e. when ACC is 0 (or below 10\%). The diagonal dashed lines indicate the day of the test year 2021 as labelled on the x-axis, the arrows indicate where forecasts reach the second test year 2022.}
    \label{fig:forecast_horizons}
\end{figure}

Forecast horizons were computed for two European regions, of which the northern one represents the area of lowest emulators’ skill (see Figure \ref{fig:europe}, b.1-3) and the southern one an area of stronger emulators’ skill. Being strongly correlated with soil water volume, these two regions differ specifically in their average snow cover fraction (see Figure \ref{fig:forecast_horizons}). The displayed horizons were computed over all prognostic state variables simultaneously, while their interpretation is related to horizons computed for prognostic state variables separately, for the figures of which we refer to the Appendix. 

In the North, predictive skill depended on an interaction of how far ahead a prediction was made (the lead time) and the day of year to which the prediction was made. In the best case, the LSTM, summer predictions were poor (light patches in Figure \ref{fig:forecast_horizons} heat maps), but only when initialised in winter. Or, in other words, one can make good predictions starting in winter, but not to summer. Vertical structures indicate a systematic model error that appears at specific initialisation times and that is independent of prediction date, for example in XGB forecasts that are initialized in May (see Figure \ref{fig:forecast_horizons}, northern region). Diagonal light structures in the heat maps indicate a temporally consistent error and can be interpreted as physical limits of system predictability, where the different initial forecast time doesn’t affect model scores.

All models show stronger limits in predictability and predictive ability in the northern European region (see Figure \ref{fig:forecast_horizons}, left column). MLP and XGB struggled with representing seasonal variation towards climatology at long lead times, while LSTM is strongly limited by a systematic error in certain regions. Initializing the forecast the 1 January 2021, MLP drops below an ACC of 80\% repeatedly from initialization on and then to an ACC below 10\% at the beginning of May. LSTMs performance is more robust in the beginning of the year but depletes strongly later to less than 10\% ACC in mid May. On the one hand, this represents two different characteristics of model errors: MLP forecasts for snow cover fraction are less than zero for some grid cells while LSTM forecasts for snow cover fraction remain falsely at very high levels for some grid cells, not predicting the snowmelt in May (see Appendix, A4.1). On the other hand, this represents a characteristic error due to change in seasonality: the snowmelt in this region in May happens abruptly and all emulators repeatedly over- or underpredict the exact date. 

\section{Discussion}

In the comparative analysis of emulation approaches for land surface forecasting, three primary models—LSTM (Long Short-Term Memory networks), MLP (Multi-Layer Perceptrons), and XGB (Extreme Gradient Boosting)—have been evaluated to understand their effectiveness across different operational scenarios. While all emulators achieved high predictive scores, models differ in their demand of computational resources \cite{cuiRealtimeRainfallrunoffPrediction2021} and each one offers unique advantages and faces distinct challenges, impacting their suitability for various forecasting tasks. With this work we want to present the first steps towards enabling quick offline experimentation on the land surface with ECMWF’s land surface scheme ECLand and decreasing computational demands, i.e.~in the coupled data assimilation. 

\subsection{Approximation of prognostic land surface states}

The total evaluation scores of our emulators indicate good agreement with ECLand simulations. Among the seven individual prognostic land surface states, emulators achieve notably different scores and in the transfer from the high-resolution continental to the low-resolution global scale, their performance ranking change. On average, neural network performances degrade towards the deeper soil layers, while XGB scores remain relatively stable. Also, the neural networks scores drop in the extrapolation from continental to global scale, while XGB scores also for this task remain constantly high.
In a way, these findings are not surprising. It is known that neural networks are highly sensitive to selection bias \cite{grinsztajnWhyTreebasedModels2022} and tuning of hyper-parameters \cite{bouthillierAccountingVarianceMachine2021}, suboptimal choices of which may destabilise variance in predictive skill. Previous and systematic comparisons of XGB and deep neural networks have demonstrated that neural networks can hardly be transferred to new data sets without performance loss \cite{shwartz-zivTabularDataDeep2021}. On tabular data, XGB still outperforms neural networks in most cases \cite{grinsztajnWhyTreebasedModels2022}, unless these models are strongly regularized \cite{kadraWelltunedSimpleNets2021}. The disadvantage of neural networks might lay in the rotational invariance of MLP-like architectures, due to which information about the data orientation gets lost, as well as in their instability regarding uninformative input features \cite{grinsztajnWhyTreebasedModels2022}. 

Inversely to expectations and preceding experiments, on the European data set relative to the two other models the LSTM scored better in the upper layer soil temperatures than in forecasting soil water volume and decreased in scores towards lower layers with slower processes. For training on observations, the decreasing LSTM predictive accuracy for soil moisture with lead time is discussed \cite{dattaMultiheadLSTMTechnique2023}, but reasons arising from the engineering side remain unclear. In an exemplary case of a single-objective, deterministic streamflow forecast, a decrease in recurrent neural network performance has been related with an increasing coefficient of variation \cite{guoAIbasedTechniquesMultistep2021}. In our European subregions, the signal-to-noise ratio of the prognostic state variables (computed as the averaged ratio of mean and standard deviation) is up to ten times higher in soil temperature than in soil water volume states (see Appendix, A2.1). While a small signal of the latter may induce instability in scores,  it does not explain the decreasing performance towards deeper soil layers with slow processes, where we expected an advantage of the long-term memory.

Stein’s paradox tells us that joint optimization may lead to better results if the target is multi-objective, but not if we are interested in single targets \cite{jamesEstimationQuadraticLoss1992, senerMultiTaskLearningMultiObjective2018}. While from a process perspective multi-objective scores are less meaningful than single ones, this is what we opted for due to efficiency. The unweighted linear loss combination might be suboptimal in finding effective parameters across all prognostic state variables \cite{chenGradNormGradientNormalization2017, senerMultiTaskLearningMultiObjective2018}, yet being strongly correlated, we deemed their manual weighting inappropriate. An alternative to this provides adaptive loss weighting with gradient normalisation \cite{chenGradNormGradientNormalization2017}.

\subsection{Evaluation in time and space}

We used aggerated MAE and RMSE accuracies as a first assessment tool to conduct model comparison, but score aggregation hides model specific spatio-temporal residual patterns. Further, both scores are variance dependent, favouring low variability in model forecasts even though this may not be representative of the system dynamic \cite{thorpeEvaluationRecentPerformance2013}. Assessing the forecast skill over time as the relative proximity to a subjectively chosen benchmark helps disentangling areas of strengths and weaknesses in forecasting with the emulators \cite{pappenbergerHowKnowIf2015}. The naïve 6-hourly climatology as benchmark highlights periods where emulators long-range forecasts on the test year are externally limited by seasonality, i.e. system predictability, and where they are internally limited by model error, i.e. the model’s predictive ability. Applying this strategy in two exemplary European subregions showed that all emulators struggle most in forecasting the period from late summer to autumn, unless they are initialized in summer (see Figure \ref{fig:forecast_horizons}). Because forecast quality is most strongly limited by snow cover (see Appendix, A4.1), we interpret this as the unpredictable start of snow fall in autumn. External predictability limitations seem to affect the LSTM overall less than the two other models, and specifically XGB drifts at long lead times.
From a geographical perspective inferred from the continental scale, emulators struggle in forecasting prognostic state variables in regions with complicated orography and strong environmental gradients. XGB scores vary seemingly random in space, while neural networks scores exhibit spatial autocorrelation. A meaningful inference about this, however, can only be conducted in assessing model sensitivities to physiographic and meteorological fields through gradients and partial dependencies. While the goal of this work is to introduce our approach to emulator development, we envision this for follow-up analyses.

\subsection{Emulation with memory mechanisms}

Without much tuning, XGB challenges both LSTM and MLP for nearly all variables (see tables 2-4). In training on observations for daily short-term and real-time rainfall-runoff prediction, XGB and LightXGB were shown before to equally performed as, or outperformed LSTMs \cite{chenGradNormGradientNormalization2017, cuiRealtimeRainfallrunoffPrediction2021}. Nevertheless, models with memory mechanism such as the encoder-decoder LSTM remain a promising approach for land surface forecasting regarding their differentiability \cite{hatfieldBuildingTangentLinear2021}, their flexible extension of lead times, for exploring the effect of long-term dependencies or for inference from the context vector that may help identifying the process relevant climate fields \cite{leesHydrologicalConceptFormation2022}. 

In our LSTM architecture, we assume that our model is well defined in that the context vector perfectly informs the hidden decoder states. If that assumption is violated, potential strategies are to create a skip-connection between context vector and forecast head, or to consider input of time-lagged variables or self-attention mechanisms \cite{chenImportanceShortLagtime2020}. With attention, the context vector becomes a weighted sum of alignments that relates neighbouring positions of a sequence, a feature that could be leveraged for forecasting quick processes such as snow cover or top-level soil water volume. 

Comparing average predictive accuracies across different training lead times indicates that training at longer lead times may enhance short-term accuracy of the LSTM at the cost of training runtime (see Appendix, A2). A superficial exploration of encoder length indicates no visible improvement on target accuracies if not a positive tendency towards shorter sequences. This needs an extended analysis for understanding, yet without a significant improvement by increased sequence length, GRU cells might provide a simplified and less parameterized alternative to LSTM cells. They were found to perform equally well on streamflow forecast performance before, while reaching higher operational speed \cite{guoAIbasedTechniquesMultistep2021}.

\subsection{Emulators in application}

LSTM networks with a decoder structure are valued for their flexible and fast lead time evaluation, which is crucial in applications where forecast intervals are not consistent. The structure of LSTM is well-suited for handling sequential data, allowing it to perform effectively over different temporal scales \cite{hochreiterLongShortTermMemory1997}. They provide access to gradients, which facilitates inference, optimization and usage for coupled data assimilation \cite{hatfieldBuildingTangentLinear2021, derosnayCoupledDataAssimilation2022}. Nevertheless, the complexity of LSTMs introduces disadvantages: Despite their high evaluation speed and accuracy under certain conditions, they require significant computational resources and long training times. They are also highly sensitive to hyperparameters, making them challenging to tune and slow to train, especially with large datasets.

MLP models stand out for their implementation, training and evaluation speed with yet rewarding accuracy, making them a favourable choice for scenarios that require rapid model deployment. They are tractable and easy to handle, with a straightforward setup that is less demanding computationally than more complex models. MLPs also allow for access to gradients, aiding in incremental improvements during training and quick inference \cite{hatfieldBuildingTangentLinear2021}. Despite these advantages, MLPs face challenges with memory scaling during training at fixed lead times, which can hinder their applicability in large-scale or high-resolution forecasting tasks.

XGB models are highly regarded for their robust performance with minimal tuning, achieving high accuracy not only in sample applications, but also in transfer to unseen problems \cite{shwartz-zivTabularDataDeep2021, grinsztajnWhyTreebasedModels2022}. Their simplicity makes them easy to handle, even for users with limited technical expertise in machine learning. However, the slow evaluation speed of XGB becomes apparent as dataset complexity and size increase. Although generally more interpretable than deep machine learning tools, XGB is not differentiable, limiting its application in coupled data assimilation \cite{hatfieldBuildingTangentLinear2021, derosnayCoupledDataAssimilation2022} even though research on differentiable trees is ongoing \cite{popovNeuralObliviousDecision2019}.

\section{Conclusion}

In conclusion, the choice between LSTM, MLP, and XGB models for land surface forecasting depends largely on the specific requirements of the application, including the need for speed, accuracy, and ease of use. Each model's computational demands, flexibility, and operational overhead must be carefully considered to optimize performance and applicability in diverse forecasting environments. When it comes to accuracy, combined model ensembles of XGB and neural networks have been shown to yield the best results \cite{shwartz-zivTabularDataDeep2021}, but accuracy alone will not determine a single best approach \cite{bouthillierAccountingVarianceMachine2021}. Our comparative assessment underscores the importance of selecting the appropriate emulation approach based on a clear understanding of each model's strengths and limitations in relation to the forecasting tasks at hand. By developing the emulators for ECMWF’s numerical land surface scheme ECLand, we path the way towards a physics-informed ML-based land surface model that on the long run can be parametrized with observations. We also provide a pretrained model suite to improve land surface forecasts and future land reanalyses. 

\section*{Code and data availability}

Code for this analysis can be found here: https://github.com/MWesselkamp/land-surface-emulation. Data is available on request. 

\section*{Author contribution}

MW, MCha, EP, FP and GB conceived the study. MW and EP conducted the analysis. MW, MCha, MK, EP discussed and took technical decisions. SB advised on process decisions. MW, MCho and FP wrote the manuscript. EP, CFD, FP reviewed the analysis and/or manuscript.

\section*{Competing interest}
The authors declare that they have no conflict of interest.

\section*{Acknowledgements}

This work profited from discussion with Linus Magnusson, Patricia de Rosnay, Sina R. K. Farhadi and Karan Ruparell and many more. MW thankfully acknowledges ECMWF for providing two research visit stipendiates over the course of the collaboration. EP was funded by the CERISE project (grant agreement No101082139) funded by the European Union. Views and opinions expressed are however those of the authors only and do not necessarily reflect those of the European Union or the Commission. Neither the European Union nor the granting authority can be held responsible for them. ChatGPT version 4.0 was used for coding support.

\bibliographystyle{unsrt}  
\bibliography{land-emulation}  

\end{document}


\maketitle
\footnotetext[4]{BrainLinks-BrainTools, Freiburg, Germany}
\footnotetext[5]{World Meteorological Organization, Geneva, Switzerland}

\section{Data base}

\subsection{European subregions for horizons computation}

A northern European subset was selected on Southern Scandinavia with a grid box on minimum and maximum latitudes of 55 and 71 degree respectively and a minimum and maximum longitude of 5 and 20 degree respectively. This resulted in a subset of 755 grid cells. For the southern European region, a grid box was created over France with minimum and maximum latitudes of 41.5 and 51.1 degree respectively and a minimum and maximum longitude of -5.1 and 6 degree respectively. Summary statics for the prognostic state variables in these regions are listed in table \ref{tab:europe_comparison}.

\begin{table}[H]
    \centering
    \renewcommand{\arraystretch}{1.5} 
    \setlength{\tabcolsep}{12pt} 
    \begin{tabular}{|c|c|c|c|c|c|c|}
        \hline
        \multirow{2}{*}{} & \multicolumn{3}{|c}{Northern Europe} & \multicolumn{3}{c|}{Southern Europe} \\ \cline{2-7} 
                          & Mean & Standard dev. & SNR & Mean & Standard dev. & SNR \\ \hline
        SWVL1 & 0.2858 & 0.0465 & 6.399 & 0.2929 & 0.0905 & 3.3495 \\ \hline
        SWVL2 & 0.2802 & 0.0433 & 6.7156 & 0.2949 & 0.0807 & 3.7471 \\ \hline
        SWVL3 & 0.2685 & 0.0449 & 6.1867 & 0.2905 & 0.0688 & 4.3294 \\ \hline
        STL1  & 278.1943 & 6.2549 & 45.6081 & 285.026 & 6.9303 & 41.67 \\ \hline
        STL2  & 278.0838 & 5.6185 & 50.9871 & 284.9675 & 6.007 & 48.0569 \\ \hline
        STL3  & 277.8869 & 4.4763 & 65.1102 & 284.847 & 4.8378 & 59.6688 \\ \hline
        SNOWC & 36.5848 & 37.9657 & 0.889 & 2.7402 & 9.5118 & 0.1722 \\ \hline
    \end{tabular}
    \caption{Exemplary summary statistics of the seven prognostic target variables over two European training data subsets, northern and southern. Mean, standard deviation and their ratio (Signal-to-noise ratio, SNR) are aggregated over times and grid cells.}
    \label{tab:europe_comparison}
\end{table}

\section{Model development}

\subsection{LSTM}

\subsubsection{Architecture and Hyperparameter selection}

\begin{figure}[H]
    \centering
    \includegraphics[width=0.9\linewidth]{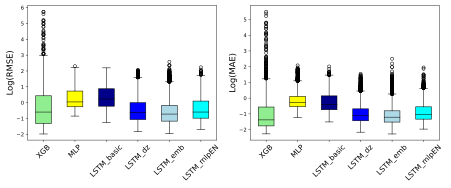}
    \caption{LSTM architecture development. LSTM\_basic considers prognostic state variables in the encoder as input, LSTM\_dz adds and incremental term in the loss function, LSTM\_emb encodes prognostic state variables to inform encoder hidden and cell states and LSTM\_mlpEN uses an MLP encoder to inform the hidden states of the LSTM decoder.}
    \label{fig:supp_fig01}
\end{figure}

The coarse architectural modules of the LSTM were manually selected. In a seeded experiment, we (1) added the first differences to the loss function, (2) added an embedding layer that transfers prognostic states to the initial hidden states of an LSTM encoder, (3) tested an MLP as encoder to the LSTM decoder (see figure \ref{fig:supp_fig01}). While we accepted the methodology of (1) and (2), we rejected (3) and continued with an LSTM encoder network.
Detaild architectural choices were made with the Bayesian hyperparameter tuning framework Optuna \cite{akibaOptunaNextgenerationHyperparameter2019}. The best performance was reached with equal parametric capacities in the encoder and decoder part. The final LSTM thus has a hidden size of 200 and in each layer with a depth of 3 in the encoder and decoder part. The parts are connected by a hidden and a cell adapter that consist each of a single linear layer that transfers the hidden and cell state from the encoder to the decoder, performing width. 
The hyperparameters for training were a dropout of 0.1265, a learning rate of  0.0005 and weight decay of 0.0001.

\paragraph{Parameters:}
\begin{itemize}
    \item hidden encoder (linear): 5.8K
    \item cell encoder (linear): 5.8K
    \item lstm encoder (LSTM): 824K
    \item hidden adapter (linear): 40.2K
    \item lstm decoder (LSTM): 40.2K
    \item hidden encoder (linear): 1.4K
\end{itemize}

\subsubsection{Training Leadtime}

\begin{figure}[!h]
    \centering
    \includegraphics[width=0.8\linewidth]{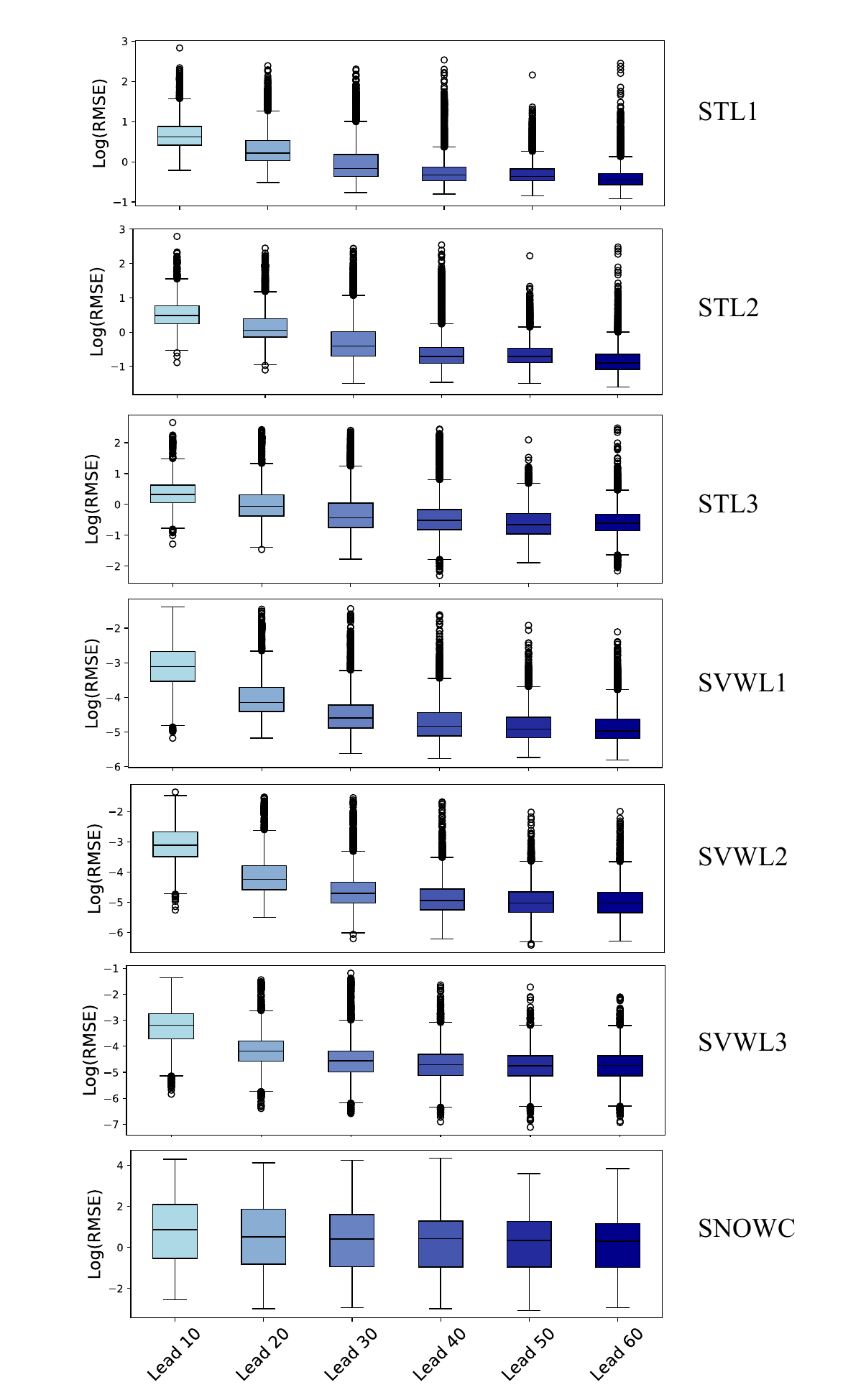}
    \caption{Targetwise predictive accuracy by training at different lead times.}
    \label{fig:supp_fig02}
\end{figure}

The forget gate mechanism allows LSTMs to store information over long time sequences without the loss old information (e.g. \cite{nearingGlobalPredictionExtreme2024}). We conducted a seeded experiment on the effect of the training lead time in the decoder part on the LSTMs predictive accuracy within the capacity of our computational resources. At the exact same hyperparameter setting, the model was trained at six different lead times for 220 epochs. Note that lead times are reported in time steps on the 6-hourly resolution, i.e. a lead time of ten is equivalent to a 2.5 days forecast, a lead time of 20 to 5 days, etc. All models converged in the training period, a gaussian kernel density estimation (KDE) of the PDF of the last 40 epochs is displayed in figure \ref{fig:supp_fig02}. The KDE was computed with a bandwidth of 0.5 and the high values of the KDE estimate indicates the small absolute variability of the losses. While predictive accuracy increases at longer training lead times, so does the training runtime (see table \ref{tab:training_leadtime}). 
The training lead time for results we show in the main manuscript was 40 on the European and 60 on the continental scale.

\begin{table}[!h]
    \centering
    \renewcommand{\arraystretch}{1.5} 
    \setlength{\tabcolsep}{8pt} 
    \begin{tabular}{|>{\centering\arraybackslash}p{1.6cm}|>{\centering\arraybackslash}p{1.6cm}|>{\centering\arraybackslash}p{1.6cm}|>{\centering\arraybackslash}p{1.6cm}|>{\centering\arraybackslash}p{1.6cm}|>{\centering\arraybackslash}p{1.6cm}|}
        \hline
        Training Leadtime & Training Runtime & Evaluation Runtime & Total RMSE & Total MAE & Total R2 \\ \hline
        10 & 420.72 & 0.016 & 1.6520 & 0.9929 & 0.9991 \\ \hline
        20 & 664.27 & 0.009 & 1.3992 & 0.8012 & 0.9991 \\ \hline
        30 & 905.47 & 0.009 & 1.1084 & 0.5958 & 0.9992 \\ \hline
        40 & 1289.27 & 0.008 & 0.9138 & 0.4983 & 0.9994 \\ \hline
        50 & 1954.62 & 0.009 & 0.7411 & 0.3691 & 0.9997 \\ \hline
        60 & 2338.86 & 0.009 & 0.6918 & 0.3459 & 0.9998 \\ \hline
    \end{tabular}
    \caption{Summary of runtimes and total mean predictive scores at different training lead times. Training was conducted on 2 GPUs, Evaluation on 1 GPU. Runtimes are reported in minutes.}
    \label{tab:training_leadtime}
\end{table}

\subsubsection{Encoder sequence length}

Like the experiment on the training lead time, we conducted a seeded experiment on the effect of encoder sequence length on predictive accuracies at a training lead time of 40. However, in contrast to varying the training lead time, changing the encoder sequence will change the model structure \cite{hochreiterLongShortTermMemory1997}. The effect not being as clear as for training lead time, we may hypothesise an advantage of shorter sequence length for the soil related variables. Models that produced results in the main manuscript were trained with encoder sequence lengths of 12 on the European and 24 on the global scale.

\begin{figure}[!h]
    \centering
    \includegraphics[width=0.8\linewidth]{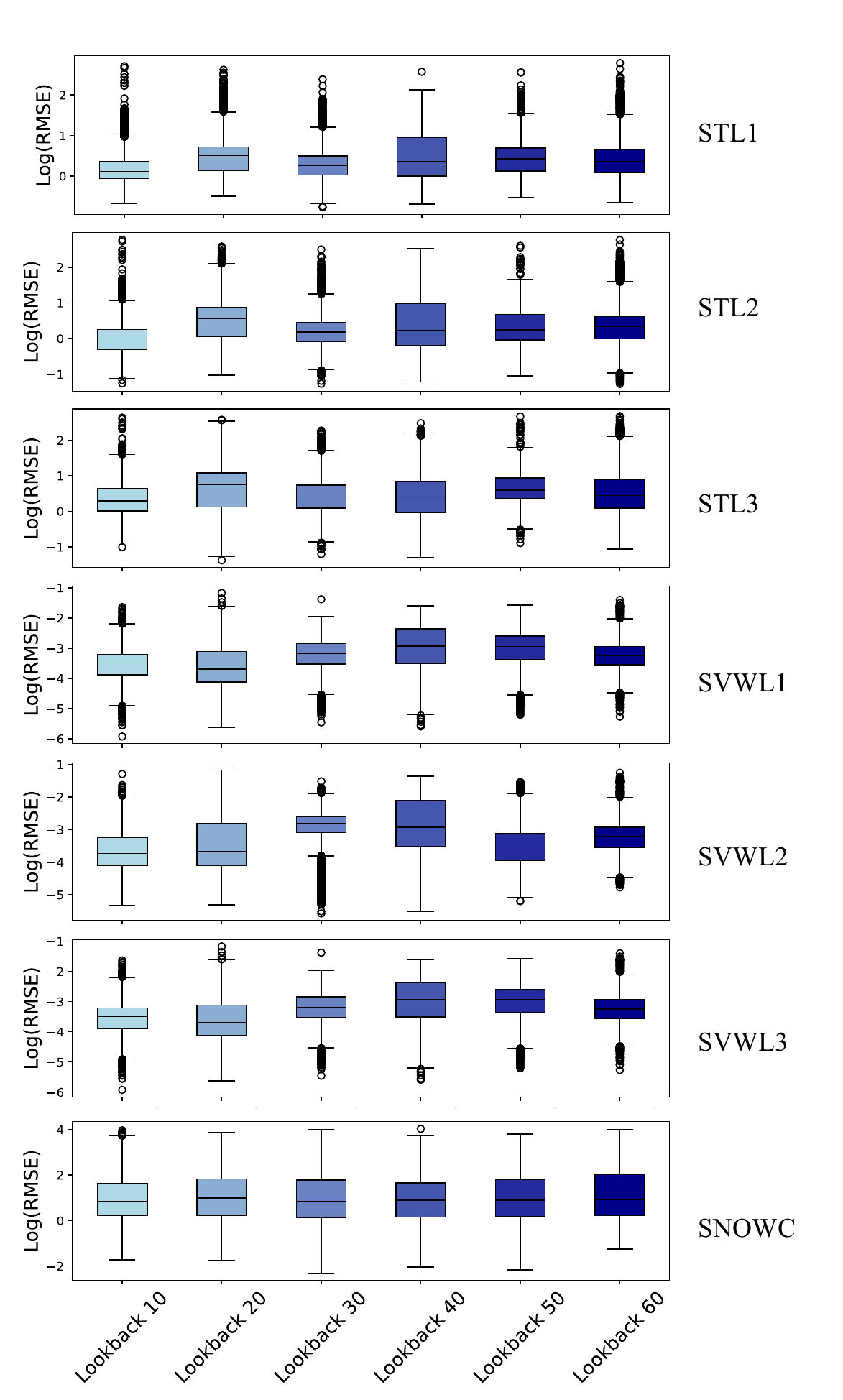}
    \caption{Targetwise predictive accuracy by training with different lookback times, i.e. encoder sequence lengths.}
    \label{fig:supp_fig03}
\end{figure}

\subsection{MLP}

\subsubsection{Methodology}

The multilayer perceptron is a neural network regression-type model that approximates a non-linear function $f: x \rightarrow y$, where $x$ in this study is a vector of static, dynamic and prognostic state variables, and $y$ the vector of prognostic state variables. The optimal function f representing this mapping is unknown and its best possible approximation $f^{*}(x)$ is found in a stochastic gradient-based optimization procedure. In practice, n non-linear functions are chained to a feed-forward neural network to create a hierarchically structured latent space with so-called hidden layers, whereby each j-th hidden layer of the network can be expressed as 
\begin{equation}
    y_j  = \phi_j (\sum_i x_i  A_{j,i}  + b_j).
\end{equation}
Here, $A_{(j,i)}$ constitutes the weight matrix, i.e. the networks parameter, $b_j$ the bias vector, i.e. an estimated intercept, and $\phi_j$ a non-linear activation function. The activation function is here the Rectified-Linear Unit (ReLU) that is defined as
$\phi = max(y,0)$. 
In a hidden layer, the input $x_i$ is mapped to a predetermined number of hidden nodes, i.e.~the layers’ size, determined by the second dimension in $A_{(j,i)}$. The transformation with $\phi_j$ returns a weighted version of the node. When weighted to zero, a node is dead unless regularized by the bias. The MLP in trained with dropout, referring to an additional regularization technique that applies a random binary mask to all input and hidden nodes of the network at each training step, where a node with the zero at the mask is dead in this training step. The probability of ones in the mask is defined as a hyperparameter (see below) \cite{goodfellowDeepLearning2016}.

\subsubsection{Architecture and Hyperparameters}

The MLP has four hidden layers of sizes 122, 47, 103 and 117. It is trained with a learning rate of 0.00093, dropout of 0.18526 and a weight decay of 0.00013. The batch size and training rollout were determined by GPU memory and are 4 and 4 respectively. The total numbers of trainable parameters in the MLP is 28.8K. 

\subsection{XGB}

XGB was trained with a learning rate of 0.3, a maximum depth of ten and 256 trees. In contrast to neural network hyperparameter optimization, only a manual exploration on tuning the learning rate and depth was conducted.

\section{Model performances}

\subsection{Model development: Europe}

\subsubsection{Objective forecast accuracies}

All emulators approximated the numerical model with high total scores on average, i.e.~R$^{2}$ values larger than 0.99, MAEs smaller than 1 and RMSE smaller than approximately 1.60. The LSTM scored highest across all metrics, followed by XGB and then MLP, even though the latter got second place in RMSE. LSTM improved in MAE by 50\% towards XGB (see table \ref{tab:metrics_all_variables}-\ref{tab:metrics_snow_cover}). These results differentiate for individual target variables. LSTM shows specifically strong performance across scores in forecasting soil water volume. 

\begin{table}[!h]
    \centering
    \renewcommand{\arraystretch}{1.5} 
    \setlength{\tabcolsep}{12pt} 
    \begin{tabular}{|>{\centering\arraybackslash}p{1.5cm}|>{\centering\arraybackslash}p{1.5cm}|>{\centering\arraybackslash}p{1.5cm}|>{\centering\arraybackslash}p{1.5cm}|>{\centering\arraybackslash}p{1.5cm}|}
        \hline
        Variable & Model & RMSE & MAE & R\textsuperscript{2} \\ \hline
        \multirow{3}{*}{All variables} & XGB & 1.6035 & 0.8091 & 0.9960 \\ \cline{2-5}
                                       & MLP & 1.6013 & 0.9611 & 0.9991 \\ \cline{2-5}
                                       & LSTM & 0.8507 & 0.4361 & 0.9996 \\ \hline
    \end{tabular}
    \caption{Model Performance Metrics for All Variables}
    \label{tab:metrics_all_variables}
\end{table}

\begin{table}[!h]
    \centering
    \renewcommand{\arraystretch}{1.5} 
    \setlength{\tabcolsep}{10pt} 
    \begin{tabular}{|>{\centering\arraybackslash}p{2.5cm}|>{\centering\arraybackslash}p{1cm}|>{\centering\arraybackslash}p{1.5cm}|>{\centering\arraybackslash}p{1.5cm}|>{\centering\arraybackslash}p{1.5cm}|>{\centering\arraybackslash}p{1.5cm}|}
        \hline
        Variable & Layer & Model & RMSE & MAE & R\textsuperscript{2} \\ \hline
        \multirow{9}{*}{Soil water volume} & \multirow{3}{*}{1} & XGB & 0.0122 & 0.0084 & 0.8442 \\ \cline{3-6}
                                           &                    & MLP & 0.0249 & 0.0192 & 0.7340 \\ \cline{3-6}
                                           &                    & LSTM & 0.0114 & 0.0083 & 0.8655 \\ \cline{2-6}
                                           & \multirow{3}{*}{2} & XGB & 0.0104 & 0.0070 & 0.8512 \\ \cline{3-6}
                                           &                    & MLP & 0.0280 & 0.0216 & 0.5781 \\ \cline{3-6}
                                           &                    & LSTM & 0.0097 & 0.0073 & 0.8543 \\ \cline{2-6}
                                           & \multirow{3}{*}{3} & XGB & 0.0149 & 0.0112 & 0.6426 \\ \cline{3-6}
                                           &                    & MLP & 0.0252 & 0.0197 & 0.2380 \\ \cline{3-6}
                                           &                    & LSTM & 0.0114 & 0.0092 & 0.7379 \\ \hline
    \end{tabular}
    \caption{Model Performance Metrics for Soil Water Volume by Layer}
    \label{tab:metrics_soil_water_volume}
\end{table}

\begin{table}[!h]
    \centering
    \renewcommand{\arraystretch}{1.5} 
    \setlength{\tabcolsep}{10pt} 
    \begin{tabular}{|>{\centering\arraybackslash}p{2.5cm}|>{\centering\arraybackslash}p{1cm}|>{\centering\arraybackslash}p{1.5cm}|>{\centering\arraybackslash}p{1.5cm}|>{\centering\arraybackslash}p{1.5cm}|>{\centering\arraybackslash}p{1.5cm}|}
        \hline
        Variable & Layer & Model & RMSE & MAE & R\textsuperscript{2} \\ \hline
        \multirow{9}{*}{Soil temperature} & \multirow{3}{*}{1} & XGB & 0.8730 & 0.5735 & 0.9750 \\ \cline{3-6}
                                          &                    & MLP & 1.2629 & 0.9601 & 0.9352 \\ \cline{3-6}
                                          &                    & LSTM & 0.8450 & 0.6347 & 0.9642 \\ \cline{2-6}
                                          & \multirow{3}{*}{2} & XGB & 0.6449 & 0.3843 & 0.9721 \\ \cline{3-6}
                                          &                    & MLP & 0.9984 & 0.7580 & 0.3130 \\ \cline{3-6}
                                          &                    & LSTM & 0.6563 & 0.4852 & 0.9480 \\ \cline{2-6}
                                          & \multirow{3}{*}{3} & XGB & 0.6221 & 0.4368 & 0.9126 \\ \cline{3-6}
                                          &                    & MLP & 1.3464 & 1.0020 & -0.5478 \\ \cline{3-6}
                                          &                    & LSTM & 0.7530 & 0.5884 & -0.5807 \\ \hline
    \end{tabular}
    \caption{Model Performance Metrics for Soil Temperature by Layer}
    \label{tab:metrics_soil_temperature}
\end{table}

\begin{table}[!h]
    \centering
    \renewcommand{\arraystretch}{1.5} 
    \setlength{\tabcolsep}{10pt} 
    \begin{tabular}{|>{\centering\arraybackslash}p{2.5cm}|>{\centering\arraybackslash}p{1cm}|>{\centering\arraybackslash}p{1.5cm}|>{\centering\arraybackslash}p{1.5cm}|>{\centering\arraybackslash}p{1.5cm}|>{\centering\arraybackslash}p{1.5cm}|}
        \hline
        Variable & Layer & Model & RMSE & MAE & R\textsuperscript{2} \\ \hline
        \multirow{3}{*}{Snow cover} & \multirow{3}{*}{top} & XGB & 9.0471 & 4.2423 & 0.5325 \\ \cline{3-6}
                                    &                      & MLP & 7.5232 & 3.9469 & 0.4383 \\ \cline{3-6}
                                    &                      & LSTM & 3.6676 & 1.3196 & 0.4345 \\ \hline
    \end{tabular}
    \caption{Model Performance Metrics for Snow Cover}
    \label{tab:metrics_snow_cover}
\end{table}

\begin{figure}[!h]
    \centering
    \includegraphics[width=0.8\linewidth]{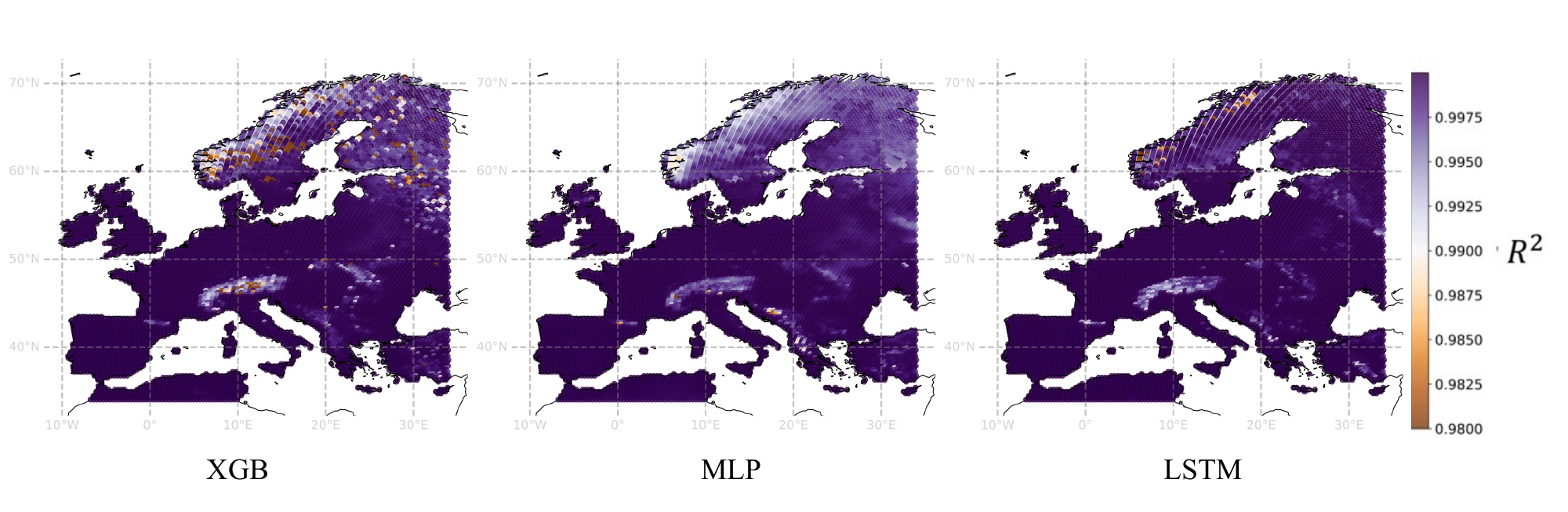}
    \caption{Mean R-squared aggregated per grid cells over 6-hourly lead times on the European subset for model development.}
    \label{fig:supp_fig05}
\end{figure}

\begin{figure}[!h]
    \centering
    \includegraphics[width=0.8\linewidth]{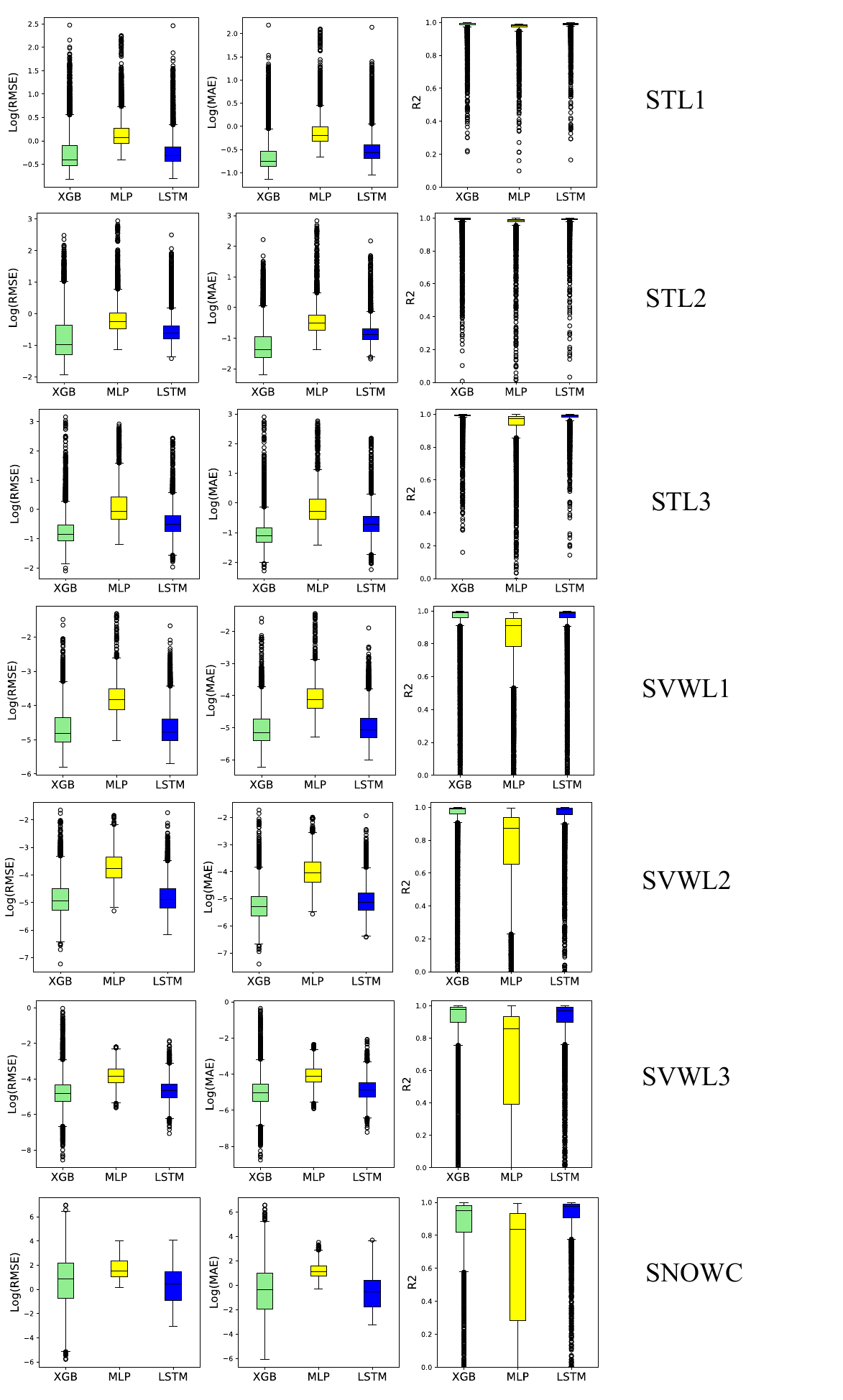}
    \caption{Total distribution of mean scores, aggregated over 6-hourly lead times by grid cell, variability here thus refers to performance differences among grid cells.}
    \label{fig:supp_fig06}
\end{figure}

\subsection{Model testing: Europe}

\subsubsection{Quantile Correlations}

\begin{figure}[!h]
    \centering
    \includegraphics[width=0.8\linewidth]{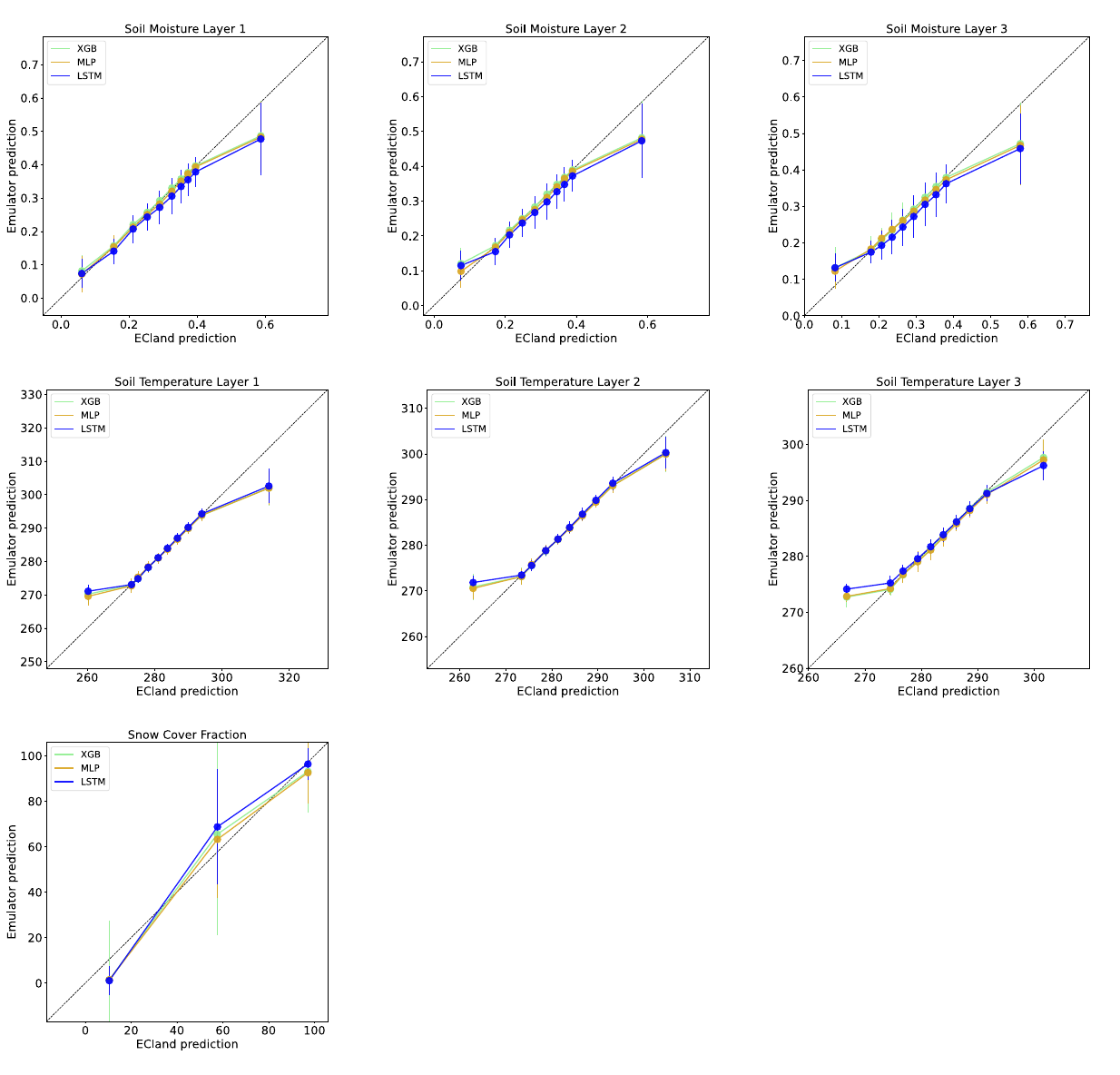}
    \caption{Quantile correlations for all prognostic target variables and all emulators. Emulator quantile predictions are on the y-axis, ECland predictions on the x-axis. The dashed black line indicates their perfect correlation.}
    \label{fig:supp_fig07}
\end{figure}

We visualised quantile correlations for each prognostic state variable. The mean and standard deviation of quantiles were computed in 10\% steps for emulator and ECland forecasts and plotted against each other. The results highlight the state values where model predictions align perfectly, i.e. quantiles are found on the correlation line, and where the emulator overestimate (quantiles above regression line) or underestimate (quantiles below regression line) ECland prognostic states (see figure \ref{fig:supp_fig07}).

\section{Evaluation}

\subsection{Forecast horizons: climatology}

\begin{figure}[!h]
    \centering
    \includegraphics[width=0.8\linewidth]{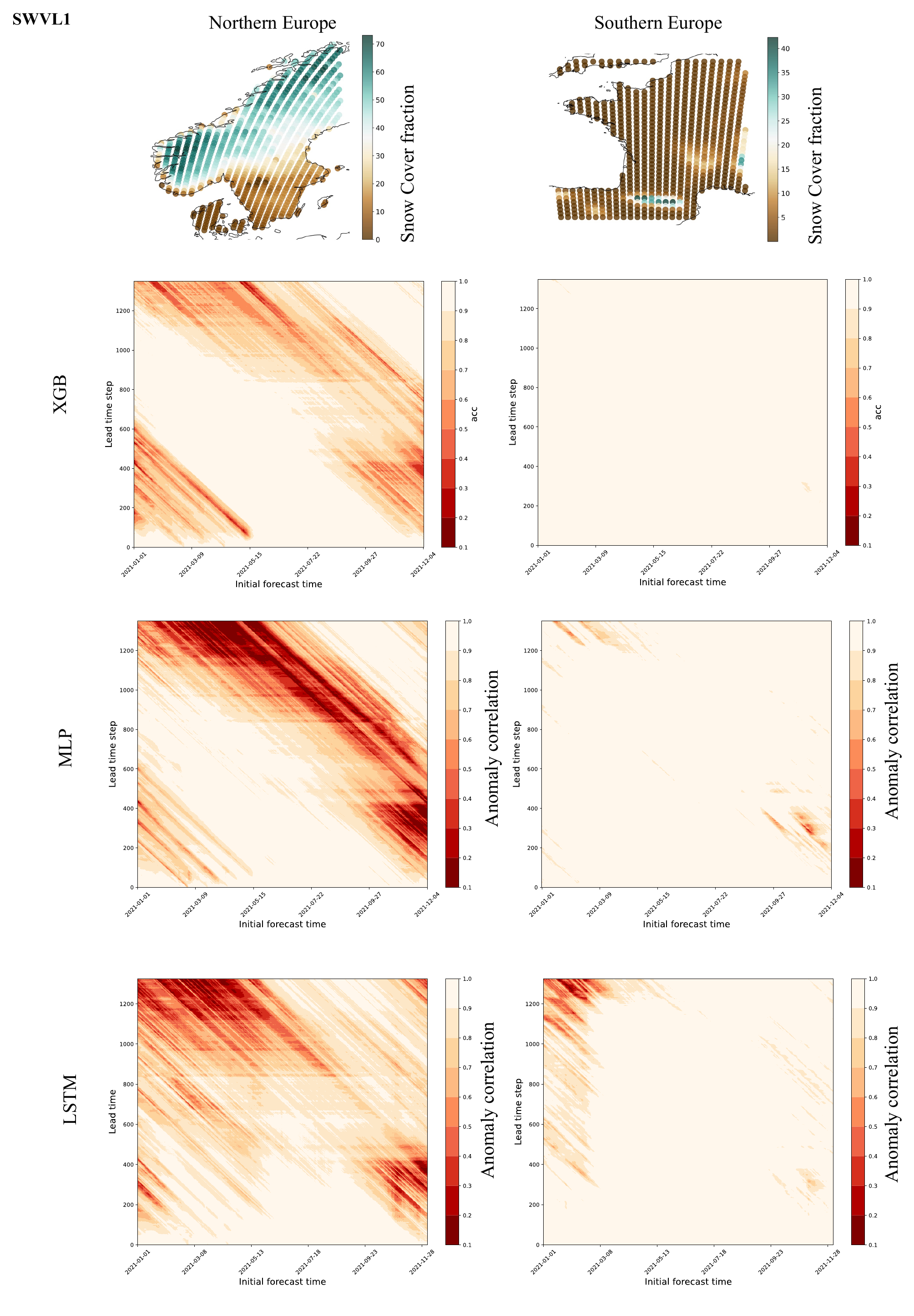}
    \caption{Forecast horizons for Soil water volume layer 1.}
    \label{fig:supp_fig08}
\end{figure}

\begin{figure}[!h]
    \centering
    \includegraphics[width=0.8\linewidth]{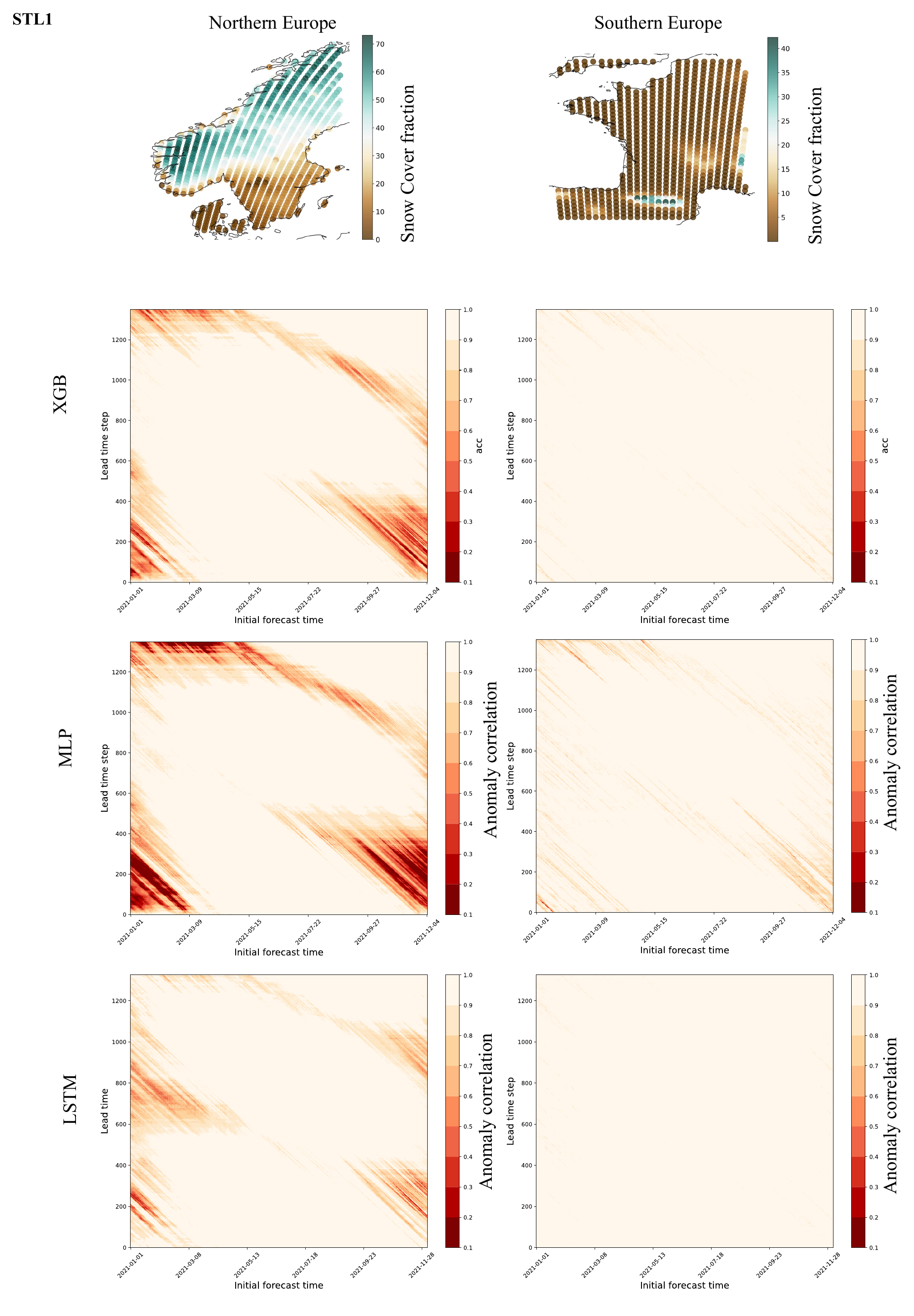}
    \caption{Forecast horizons soil temperature Layer 1.}
    \label{fig:supp_fig09}
\end{figure}

\begin{figure}[!h]
    \centering
    \includegraphics[width=0.8\linewidth]{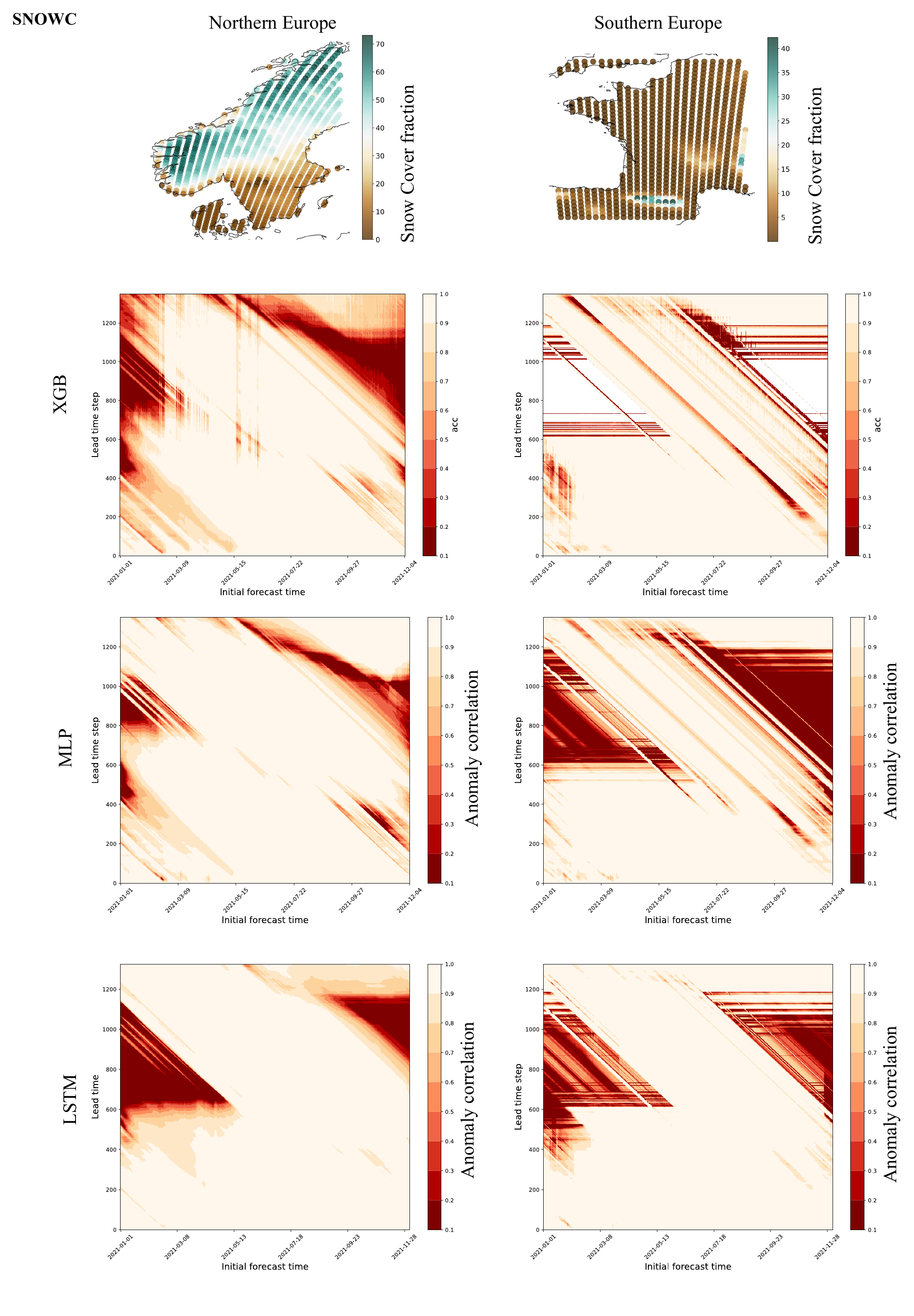}
    \caption{Forecast horizons snow cover.}
    \label{fig:supp_fig10}
\end{figure}

Below we show examples of forecast horizons computed for three single prognostic state variables, soil water volume and temperature at layer one and snow cover (figures \ref{fig:supp_fig08}-\ref{fig:supp_fig10}). Disentangling these highlights at the example of snow cover that in aggregating the anomaly correlation over prognostic state variables, negative and positive effects may cancel each other out: the snow cover limitation in the southern European subregion for the MLP forecasts is not as visible in the total horizons (see main manuscript). 

\subsection{Time series sample: Northern Europe}

\begin{figure}[!h]
    \centering
    \includegraphics[width=0.8\linewidth]{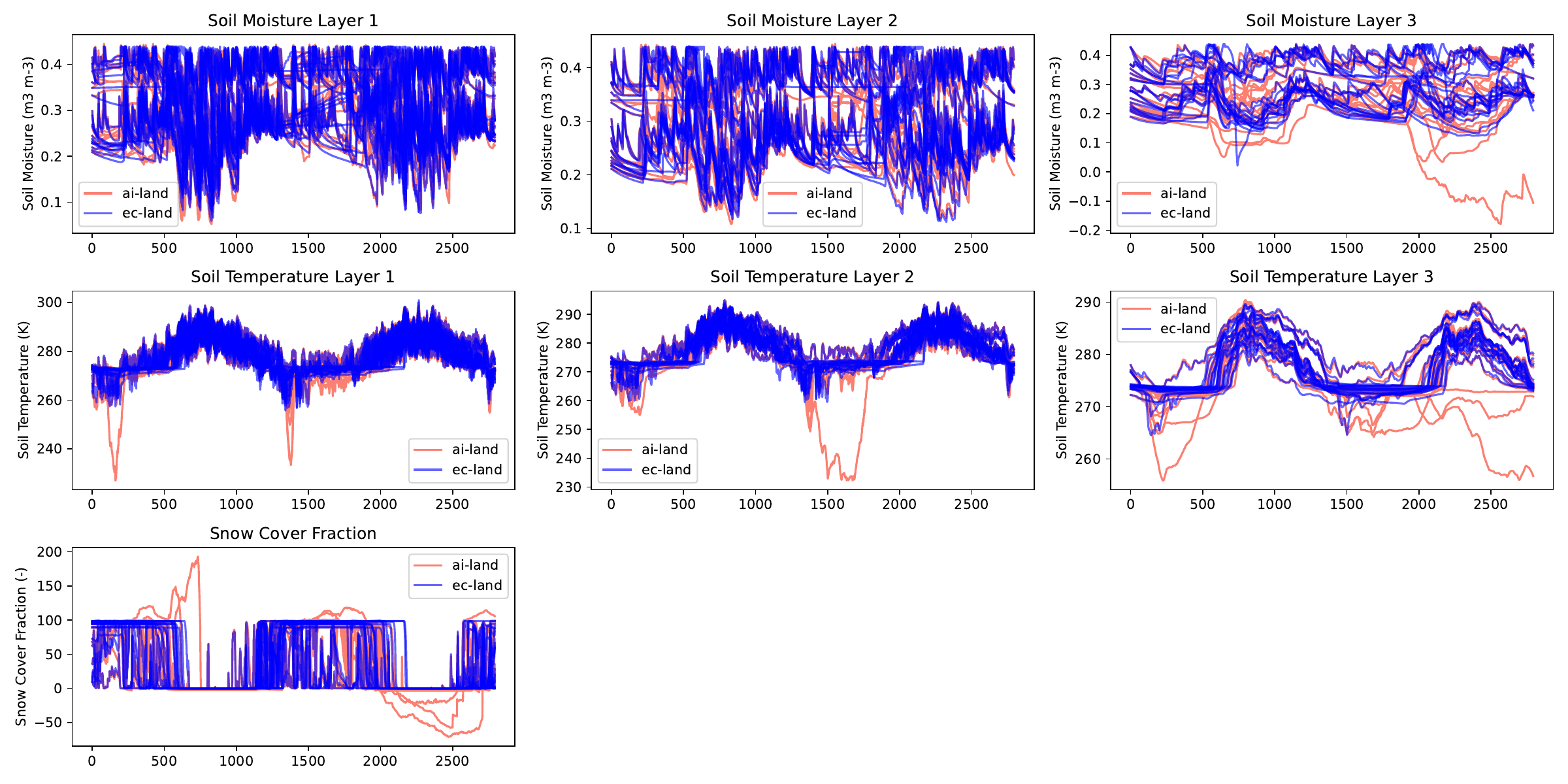}
    \caption{MLP forecast on two  test years 2021, 2022  for a random selection of grid cells from the northern European region.}
    \label{fig:supp_fig011}
\end{figure}

\begin{figure}[!h]
    \centering
    \includegraphics[width=0.8\linewidth]{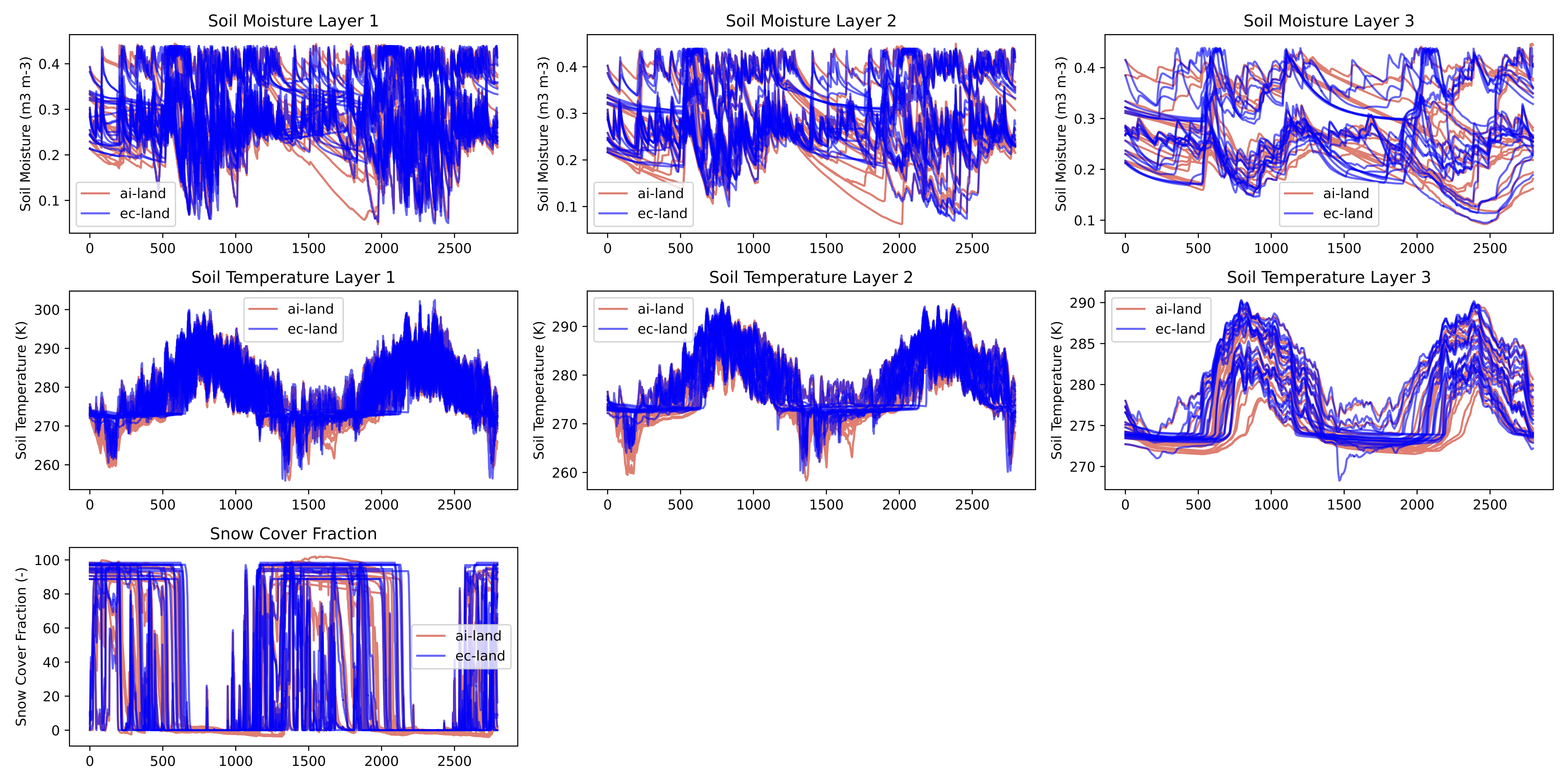}
    \caption{LSTM forecast on two test years 2021, 2022 for a random selection of grid cells from the northern European region.}
    \label{fig:supp_fig012}
\end{figure}

\begin{figure}[!h]
    \centering
    \includegraphics[width=0.8\linewidth]{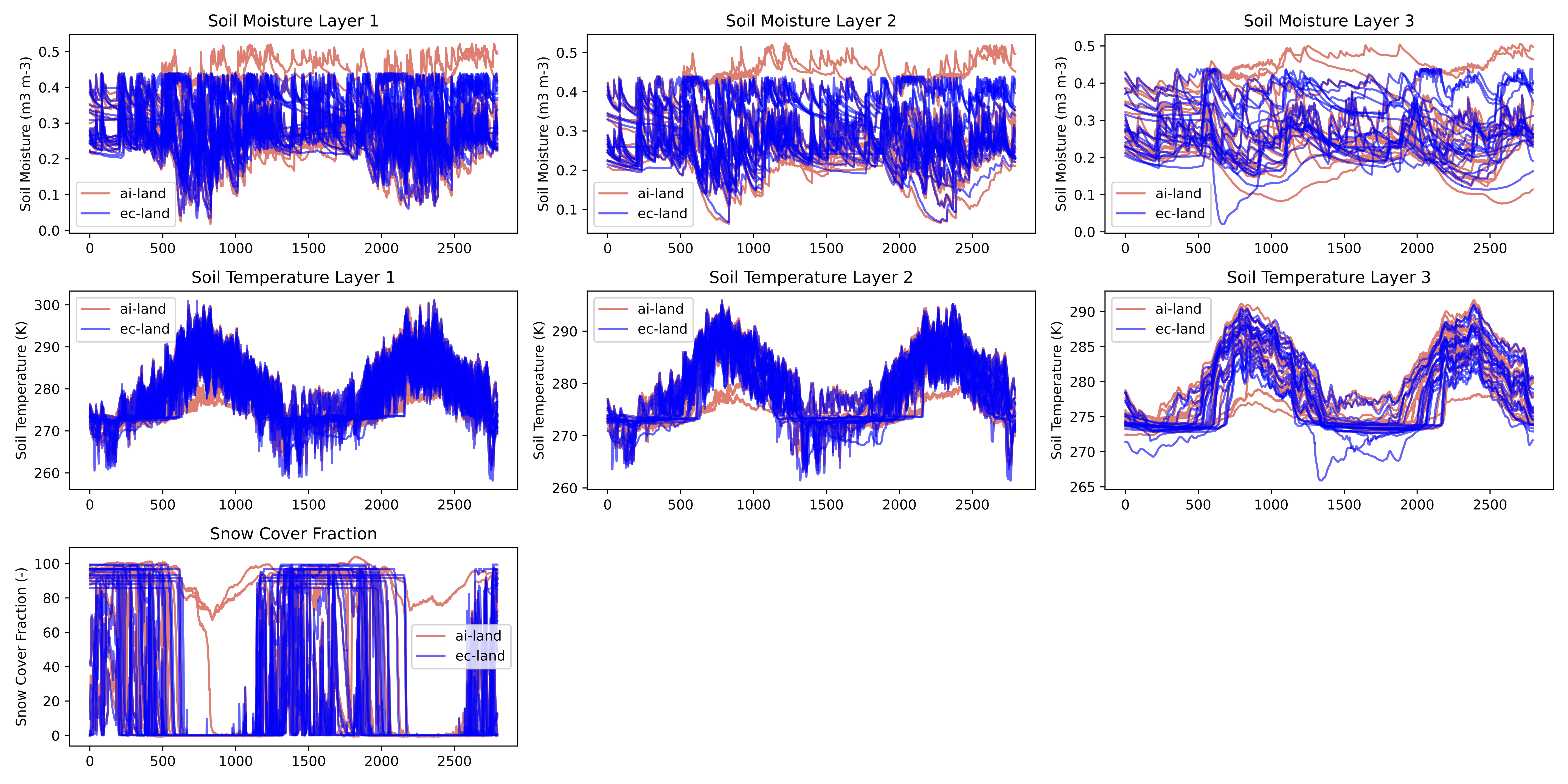}
    \caption{XGB forecast on two test years 2021, 2022  for a random selection of grid cells from the northern European region.}
    \label{fig:supp_fig013}
\end{figure}

\subsection{Model extrapolation: Globe, low-resolution (TCO199)}

\begin{figure}[!h]
    \centering
    \includegraphics[width=0.8\linewidth]{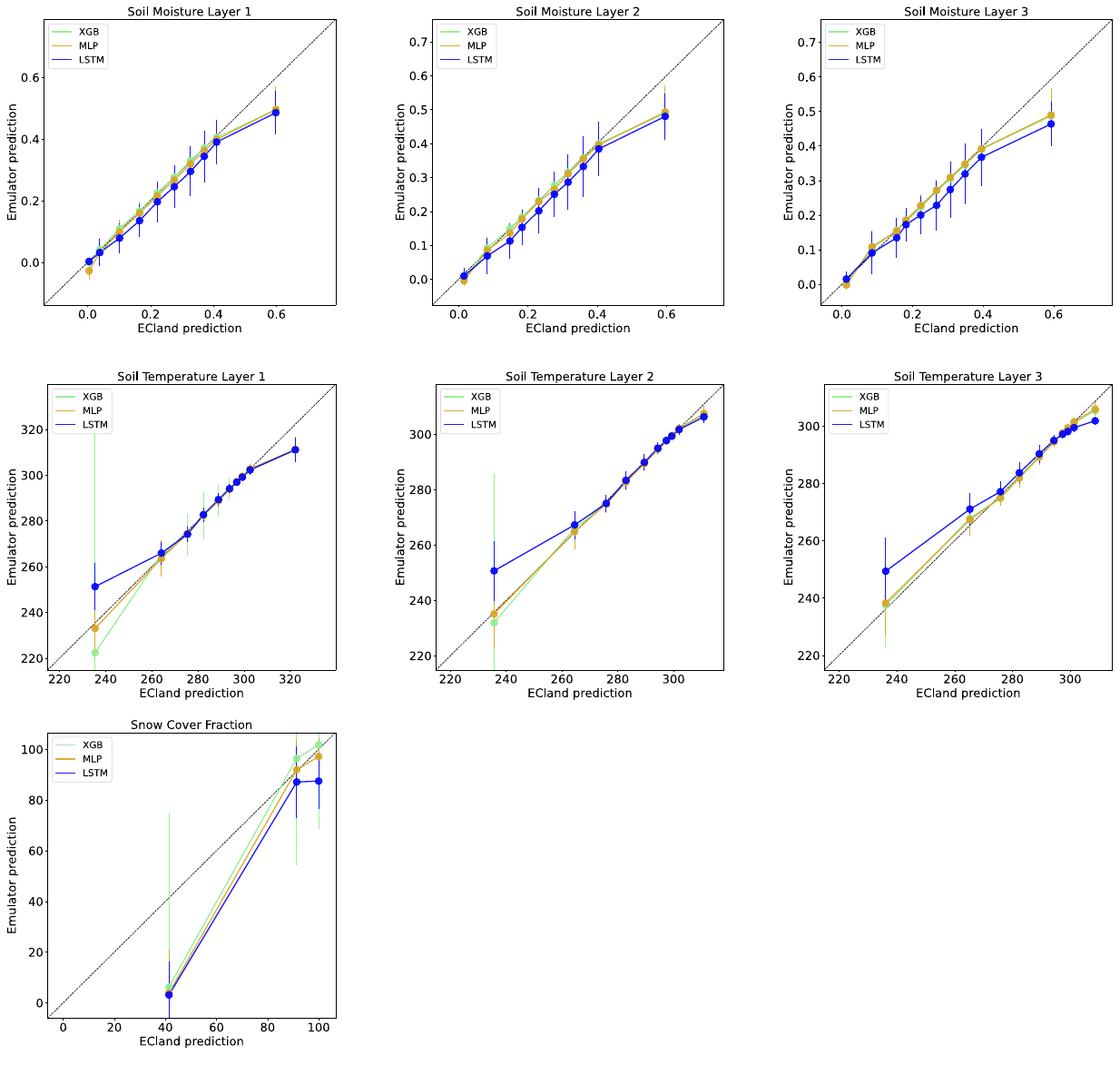}
    \caption{Quantile correlations, visualised as described in section 3.2.1 for continental model testing.}
    \label{fig:supp_fig014}
\end{figure}

\begin{figure}[!h]
    \centering
    \includegraphics[width=0.8\linewidth]{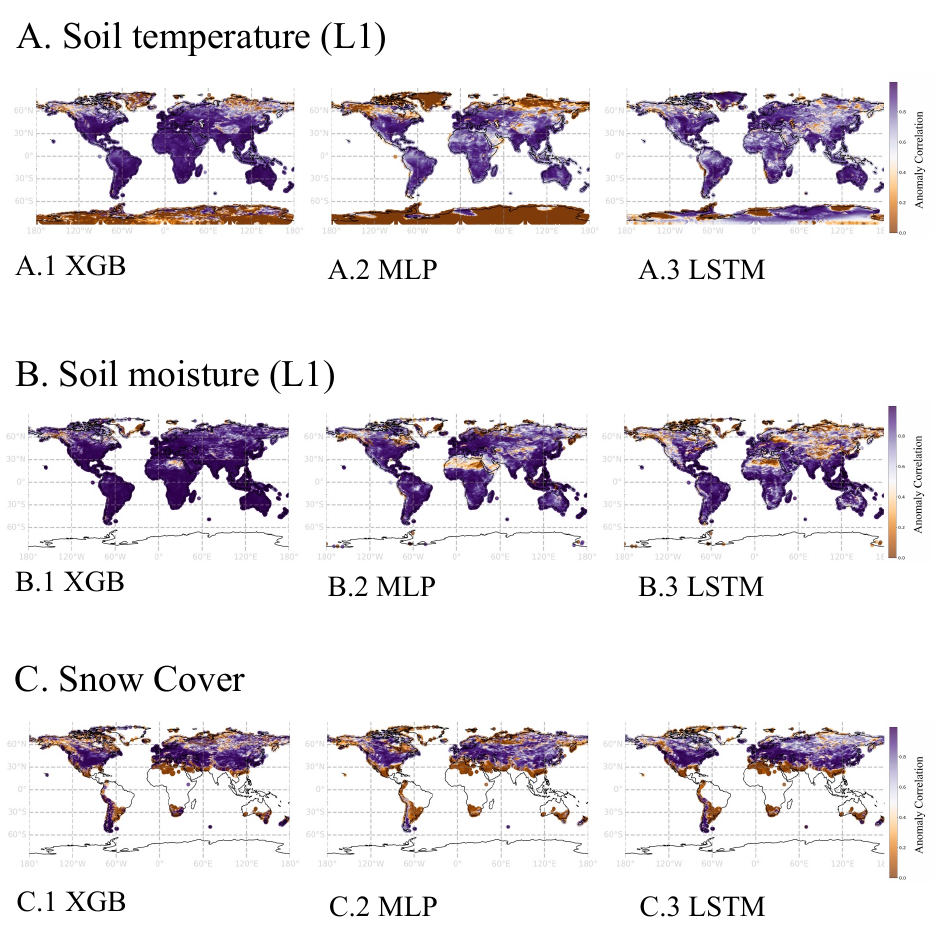}
    \caption{Global distribution of Anomaly Correlation for three prognostic state variables. Uncoloured areas indicate regions where the ACC is not defined.}
    \label{fig:supp_fig015}
\end{figure}

\bibliographystyle{unsrt}  
\bibliography{land-emulation}  